\ifdefined\pdfoutput\pdfoutput=1\fi
\PassOptionsToPackage{unicode}{hyperref}
\PassOptionsToPackage{hyphens}{url}
\PassOptionsToPackage{dvipsnames,svgnames,x11names}{xcolor}
\documentclass[
]{article}
\usepackage{xcolor}
\usepackage[margin=1in]{geometry}
\usepackage{amsmath,amssymb}
\usepackage{iftex}
\ifPDFTeX
  \usepackage[T1]{fontenc}
  \usepackage[utf8]{inputenc}
  \usepackage{textcomp} 
  \DeclareUnicodeCharacter{0394}{\ensuremath{\Delta}}
  \DeclareUnicodeCharacter{00D7}{\ensuremath{\times}}
  \DeclareUnicodeCharacter{2026}{\ldots}
\else 
  \usepackage{unicode-math} 
  \defaultfontfeatures{Scale=MatchLowercase}
  \defaultfontfeatures[\rmfamily]{Ligatures=TeX,Scale=1}
\fi
\usepackage{lmodern}
\ifPDFTeX\else
\fi
\IfFileExists{upquote.sty}{\usepackage{upquote}}{}
\IfFileExists{microtype.sty}{
  \usepackage[]{microtype}
  \UseMicrotypeSet[protrusion]{basicmath} 
}{}
\makeatletter
\@ifundefined{KOMAClassName}{
  \IfFileExists{parskip.sty}{%
    \usepackage{parskip}
  }{
    \setlength{\parindent}{0pt}
    \setlength{\parskip}{6pt plus 2pt minus 1pt}}
}{
  \KOMAoptions{parskip=half}}
\makeatother
\usepackage{longtable,booktabs,array}
\usepackage{calc} 
\usepackage{etoolbox}
\makeatletter
\patchcmd\longtable{\par}{\if@noskipsec\mbox{}\fi\par}{}{}
\makeatother
\IfFileExists{footnotehyper.sty}{\usepackage{footnotehyper}}{\usepackage{footnote}}
\makesavenoteenv{longtable}
\usepackage{graphicx}
\makeatletter
\newsavebox\pandoc@box
\newcommand*\pandocbounded[1]{
  \sbox\pandoc@box{#1}%
  \Gscale@div\@tempa{\textheight}{\dimexpr\ht\pandoc@box+\dp\pandoc@box\relax}%
  \Gscale@div\@tempb{\linewidth}{\wd\pandoc@box}%
  \ifdim\@tempb\p@<\@tempa\p@\let\@tempa\@tempb\fi
  \ifdim\@tempa\p@<\p@\scalebox{\@tempa}{\usebox\pandoc@box}%
  \else\usebox{\pandoc@box}%
  \fi%
}
\def\fps@figure{htbp}
\makeatother
\NewDocumentCommand\citeproctext{}{}

\makeatletter
 \let\@cite@ofmt\@firstofone
 \def\@biblabel#1{}
 \def\@cite#1#2{{#1\if@tempswa , #2\fi}}
\makeatother
\newlength{\cslhangindent}
\newlength{\csllabelwidth}
\newenvironment{CSLReferences}[2] 
 {\begin{list}{}{%
  \setlength{\itemindent}{0pt}
  \setlength{\leftmargin}{0pt}
  \setlength{\parsep}{0pt}
  \ifodd #1
   \setlength{\leftmargin}{\cslhangindent}
   \setlength{\itemindent}{-1\cslhangindent}
  \fi
  \setlength{\itemsep}{#2\baselineskip}}}
 {\end{list}}
\usepackage{calc}

\providecommand{\tightlist}{%
  \setlength{\itemsep}{0pt}\setlength{\parskip}{0pt}}
\usepackage{booktabs}
\usepackage{float}
\usepackage{fvextra}
\usepackage{booktabs}
\usepackage{longtable}
\usepackage{array}
\usepackage{multirow}
\usepackage{wrapfig}
\usepackage{float}
\usepackage{colortbl}
\usepackage{pdflscape}
\usepackage{tabularx}
\usepackage{xltabular}
\usepackage{threeparttable}
\usepackage{threeparttablex}
\usepackage[normalem]{ulem}
\usepackage{makecell}
\usepackage{xcolor}
\usepackage{bookmark}
\IfFileExists{xurl.sty}{\usepackage{xurl}}{} 
\hypersetup{
  pdftitle={Beyond Gentrification: quantifying the economic{,} social and cultural dimensions of neighbourhood change through housing markets and digital traces},
  pdfauthor={A Comber1*, M Asher1, Y Wang1, M Kieu2, BT Quang3, HNT Thuy4, PH Huu5, N Malleson1,6},
  colorlinks=true,
  linkcolor={black},
  filecolor={Maroon},
  citecolor={Blue},
  urlcolor={blue},
  pdfcreator={LaTeX via pandoc}}

\title{Beyond Gentrification: quantifying the economic, social and
cultural dimensions of neighbourhood change through housing markets and
digital traces}
\author{A Comber\textsuperscript{1}*, M Asher\textsuperscript{1}, Y
Wang\textsuperscript{1}, M Kieu\textsuperscript{2}, BT
Quang\textsuperscript{3}, HNT Thuy\textsuperscript{4}, PH
Huu\textsuperscript{5}, N Malleson\textsuperscript{1,6}}
\date{November 2026}

\begin{document}
\maketitle

\begin{center}
\textsuperscript{1}School of Geography, University of Leeds, UK\\
\textsuperscript{2}Department of Civil and Environmental Engineering, University of Auckland, New Zealand\\
\textsuperscript{3}Faculty of Geography, VNU University of Science, Hanoi, Vietnam\\
\textsuperscript{4}VNU Vietnam Japan University, Hanoi, Vietnam\\
\textsuperscript{5}R\&D Consultants, Hanoi, Vietnam\\
\textsuperscript{6}Leeds Institute for Data Analytics, University of Leeds, UK\\

*email: a.comber@leeds.ac.uk\\

\end{center}

\tableofcontents

\section*{Abstract}\label{abstract}
\addcontentsline{toc}{section}{Abstract}

Neighbourhood change is a multidimensional process in which economic,
demographic, social and cultural transformations do not necessarily
coincide over space and time. However, data driven approaches typically
collapse these into single indices or regression models, losing the
nuances of individual components. This paper integrates housing market,
demographic and social media data (SMD) to characterise different
configurations of neighbourhood change in Leeds and Bradford, UK. Rates
of change (2014--2019) were estimated over Lower Super Output Areas for
house prices relative to the 250 nearest neighbouring areas, and for
residential mobility, ethnic composition, unemployment and professional
occupation. Approximately 1.6 million geo-tagged tweets, filtered for
personal users using a large language model (LLM), were analysed using
Structural Topic Models with socio-economic prevalence covariates, LLM
based neighbourhood characterisation and text embeddings scored against
gentrification and decline archetypes. These measures were used as
inputs to a model-based cluster analysis, which identified six distinct
neighbourhood types, and a spatially varying multinomial GAM was used to
examine the effects of changes in the inputs on cluster membership. The
results show repeated divergence between structural and discursive
trajectories: rapid turnover that is invisible in local SMD discourse,
heritage villages experiencing amenity and incomer pressure, a settled
British-Pakistani community whose stability would be mis-characterised
as stagnation or decline, and marginal areas accumulating strain with
little on-line signal. This suggests that SMD derived measures can
provide independent information on neighbourhood character rather than
simply confirming structural indicators, although low SMD signals may
reflect under-detection rather than an absence of change.

\section{Introduction}\label{introduction}

Neighbourhood change is a multidimensional process involving economic,
demographic, social and cultural transformations that do not necessarily
coincide over space and time. This is because the drivers of
neighbourhood transformations such as gentrification are heterogeneous
and difficult to detect and characterise at scale (Delmelle 2022; Ling
and Delmelle 2016; Modai-Snir and Ham 2019; Hochstenbach and Van Gent
2015; Yee and Dennett 2022). Individual areas may even have unique
change related signatures. This paper presents a data driven approach to
understand the impacts of changes in individual factors and how and
where they are associated with neighbourhood change. Specifically, the
analysis proposes an approach to determine the extent to which
neighbourhoods exhibit distinctive configurations and traces of social,
demographic and cultural change, rather than following a single
continuum of neighbourhood gentrification, improvement or decline.

Data driven approaches for examining neighbourhood change and
gentrification are typically based on statistical analyses of
demographic, socio-economic and housing-market indicators, commonly
derived from census and other area-based statistics, population surveys,
and property sales and rental data (Gray et al. 2021, 2023a; Reades et
al. 2023, 2019; Yee and Dennett 2022; Finio 2022; Easton et al. 2020;
Bunten et al. 2024). These have been used to model locales of
gentrification and other neighbourhood transformations through the
construction of indices, inferential and predictive regression and
classification.

Index based approaches (e.g. Flinders and Adam (2025); Almeida (2021))
typically construct a gentrification index from socio-economic
variables. These are collapsed to a single gentrification or
neighbourhood change score. Regardless of the number of inputs or the
relative weightings used in index construction, it is evident that
nuances of the index components are lost in an aggregated index. This
results in similar index scores composed of very different input values,
and neighbourhoods with similar scores having very different demographic
and structural characteristics.

In regression based approaches (e.g. Bunten et al. (2024); Hochstenbach
and Van Gent (2015); Clark and Coulter (2015)) the objective is often to
analyse some characteristic or indicator of neighbourhood change as a
response variable. The resulting coefficient estimates describe the
relationship between the input factors and the response and the model
can be used for prediction over unsampled areas or for inference
(process understanding). By definition, a good regression model is one
that captures the structure in the data via the predictor-to-response
relationships, which may be linear, exponential, quadratic etc., and
specified as such in the model. Standard regression approaches (like
ordinary least squares) create a single model to quantify spatially
stationary predictor-to-response relationships. Spatial approaches like
geographically weighted regression (GWR) (Brunsdon et al. 1996) and
spatial GAMs (Comber, Harris, and Brunsdon 2024) seek to capture spatial
non-stationarity in relationships by constructing and combining a series
of local models in GWR, or through spatial smooths in GAMs. These allow
predictor-to-response relationships to vary over space.

In both index and regression based models, the data inputs need to have
sufficient global or local structure (e.g.~the data contain meaningful
interactions between the variables) for the models to have inferential
power. This is generally the case when the variables used in the model
describe socio-economic factors such as those proposed by Gray et al.
(2023a) and Gray et al. (2023b) (House price, Professional occupation,
Residential mobility (i.e.~in-and out-migration) and Ethnic
composition), and the data are readily incorporated within standard
index and regression based approaches because the data generally move in
the same direction (i.e.~are correlated to some degree). However, they
do not capture how neighbourhoods are perceived through residents'
everyday experiences and their digitally expressed discourses.

The overarching aim of this study was to use geo-tagged social media
data (SMD) to complement socio-economic data in order to characterise
neighbourhood dynamics. SMD captures user perceptions, preferences,
attitudes, sentiments, interests, etc., and as such it can potentially
be used as a lens to understand more nuanced aspects of neighbourhood
change than are captured by standard socio-economic data. Neighbourhood
changes such as gentrification have been found to be closely linked to
social media usage (Gibbons et al. 2018; Walters and Smith 2024;
Bronsvoort and Uitermark 2022), although with different manifestations,
due to high social media engagement amongst individuals associated with
gentrification. It is used to affirm neighbourhood identity status, to
promote shifting consumption patterns within changing neighbourhoods, as
part of place co-creation (Chang and Spierings 2023; Friesenecker and
Lagendijk 2021), and to reinforce emerging neighbourhood related
processes, particularly among gentrifiers and neighbourhood neo-pioneers
(Friesenecker and Lagendijk 2021). As a result, analyses of SMD have
been used to capture emerging, local social phenomena (Lansley and
Longley 2016), to explore the changing character of neighbourhoods and
their social processes (Poorthuis et al. 2022; Bronsvoort and Uitermark
2022) and to quantify behaviours and social dynamics over neighbourhoods
(Nguyen et al. 2016). However, SMD may be sparse, noisy and uncorrelated
with socio-economic data. For these reasons, this paper uses both forms
of data and their trajectories over time and space to construct
statistical clusters of neighbourhoods, with similar socio-economic
trajectories and social media traces. It avoids the creation of indices
and regression models as SMD are highly unstructured, sparse, noisy
random and with seemingly little relationship to socio-economic data and
the mismatches between the multi-dimensional nature of neighbourhood
change and classic data driven approaches to characterise it.

The analysis investigates whether and how neighbourhood change is
reflected in the relationships --- or lack of relationships --- between
changing demographics, changing housing markets and changing social and
cultural representations of place. It demonstrates how neighbourhood
change can be quantitatively understood as a multidimensional process.
It illustrates how demographic change, economic changes in the housing
market and in the social/cultural meanings, as captured in SMD
reflecting activities associated with place, do not necessarily move
together. It does this by integrating analyses of 1) how neighbourhood
housing markets change differently from surrounding areas, 2) time
series small area demographic data that captures changes in residential
mobility, unemployment, ethnic composition and people in professional
occupations, and 3) social media data (SMD) traces through topic
modelling, large language model (LLM) interpretations and text
embeddings. The combination of these allows integrated investigations of
the relationships between facets associated with neighbourhood
transformations and changes in the economic, social, cultural and
symbolic character of those places. It examines whether some of the less
tangible dimensions of neighbourhood character can be detected
systematically from large scale digital traces, and how they relate
spatially to more conventional socio-economic indicators. These
variables, along with classic socio-economic variables, are used as
inputs to a cluster analysis to characterise different types and locales
of neighbourhood change and then used to construct a spatially varying
coefficient model that estimates the effects on neighbourhood cluster
membership arising from changes in the input variables.

By combining analyses of different neighbourhood related processes over
time, this paper addresses some of the well recognised issues associated
with disentangling different manifestations and features of
neighbourhood change at scale (Yee and Dennett 2022). It explores what
is not captured by classic socio-economic analyses by interrogating SMD
traces that implicitly relate to social activity, cultural identity and
discourse - the production and representation of neighbourhood or place
identities (Bronsvoort and Uitermark 2022). This mirrors some of the
concepts in Status--Quality--Trade-Off theory (Huu Phe and Wakely 2000)
which posits that conventional housing models capture measurable utility
and property attributes, but neighbourhood desirability also relates to
intangible place characteristics and perceptions, social and cultural
representations, as well as activities. The time series data are brought
together over neighbourhood areas and clustered to investigate the
multivariate structure of neighbourhoods and the features that
differentiate them. The analysis examines whether neighbourhoods exhibit
distinctive configurations and trajectories of social, cultural and
demographic change, rather than following a single continuum of, for
example, gentrification related upgrading or decline. This resonates
with much of the literature on heterogeneous forms of gentrification and
neighbourhood change (Butler and Robson 2001; Lees et al. 2010) and the
distinctive nature of neighbourhoods via their different combinations of
social, cultural and economic capital. It extends recent research which
has demonstrated how data-driven approaches can be used to identify and
characterise changing neighbourhoods (Gray et al. 2023b).

In this way, the front-to-back statistical analysis presented in the
paper connects to arguments in urban geography research that
neighbourhood change is simultaneously economic, social and cultural
(Cheshire et al. 2019), and to research into neighbourhood cultural
capital in which the cultural practices and identities associated with
places are implicated in change processes (Bridge 2006).

\section{Data, methods and conceptual
approaches}\label{data-methods-and-conceptual-approaches}

\subsection{House price and demographic
data}\label{house-price-and-demographic-data}

Individual house sales transactions data was summarised annually (2014
to 2019) over neighbourhood areas. This was done nationally (in order to
support a spatial analysis of neighbours) and then the data for the
local authority districts of Leeds and Bradford in West Yorkshire, UK
were extracted. A dataset of approximately 5.8 million house sales
transactions (2014--2019) in England and Wales from WhenFresh / Zoopla
was obtained via the CDRC (\url{https://data.cdrc.ac.uk}). Each
observation was geolocated using the property postcode. This allowed
each transaction to be mapped to a specific neighbourhood. In this
study, 2011 Lower Super Output Areas (LSOAs) were used as the
neighbourhood units. These represent the second finest resolution at
which census data are reported, with each LSOA typically containing
around 1,500 people (\textasciitilde500 households). All data were
subsequently summarised over these areas. Time series data describing
annual LSOA proportions of residential mobility, non-white residents,
unemployment, professional occupation as described in Gray et al.
(2023a) and Gray et al. (2023b) was used to quantify demographic change.
These variables were chosen because they describe the underpinning
dimensions of neighbourhood change (Gray et al. 2021).

\begin{figure}

{\centering \includegraphics{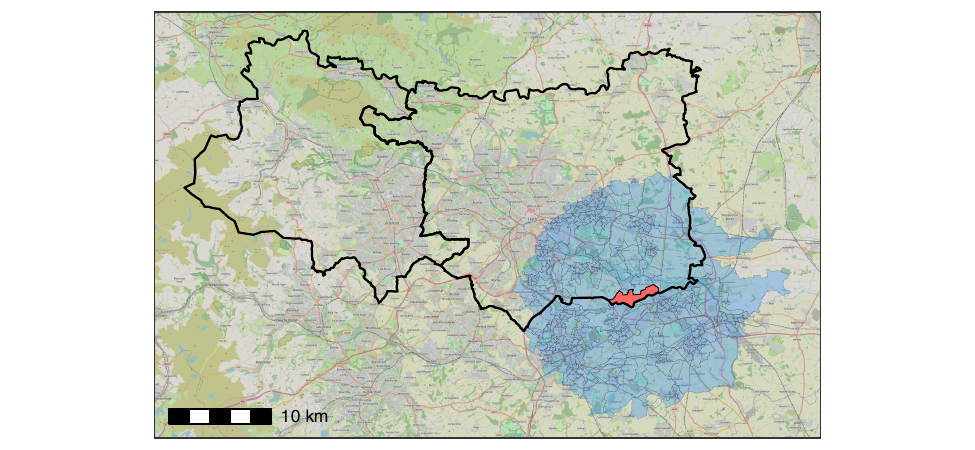} 

}

\caption{An LSOA in red (around 500 households) in the Leeds case study area and its 250 nearest neighbouring LSOA areas, with the Leeds and Bradford areas and an OpenStreetMap backdrop.}\label{fig:fig_nb}
\end{figure}

The house price data were summarised over the LSOAs nationally and
ratios over time were constructed of the average (median) price in each
LSOA to those of neighbouring or nearby LSOAs. These ratios were
regressed against time as described below. The coefficient estimate for
time could then be used to identify areas whose house prices had
increased or decreased at much greater rates than those of nearby areas,
indicating some kind of change of character of neighbourhood and
potentially suggesting areas of gentrification or decline. Formally, for
each LSOA, a ratio, \(R_t\) of the median house price, for time period,
\(i\) was calculated and compared to the average house price of nearby
areas:

\begin{equation}
R_t = \frac{ \operatorname{average}(HP_{A}) } { \operatorname{average}(HP_{NB}) }
\label{eqnRt}
\end{equation}

where, \(R_t\) is the ratio for time period \(t\),
\(\operatorname{average}(HP_{A})\) is the median price for the area
being evaluated at time \(i\) and \(\operatorname{average}(HP_{NB})\) is
the median house price in the 250 nearest LSOAs in year \(i\). An
example of the nearest-neighbours is shown in Figure \ref{fig:fig_nb}.
This approach is more robust than a \(\beta_t\) from LSOA house prices
over time because it accounts for regional economic trends, explicitly
incorporating neighbourhood-to-neighbourhood comparison and isolating
local anomalies.

Thus, 6 annual median price ratios were calculated for each LSOA, and
the values (\(V\)) were regressed against time:

\begin{equation}
V = \beta_0 + \beta_t \cdot t + \epsilon
\label{eqnBetat}
\end{equation}

where \(\beta_0\) is the intercept, \(\beta_t\) is the coefficient
estimate for time \(t\) for variable \(V\) and \(\epsilon\) is the
residual error term. As well as being generated for the house price
data, rates of change over time (\(\beta_t\)) were estimated for the 4
demographic factors of residential mobility, non-white residents,
unemployment and professional occupation.

\begin{figure}

{\centering \includegraphics{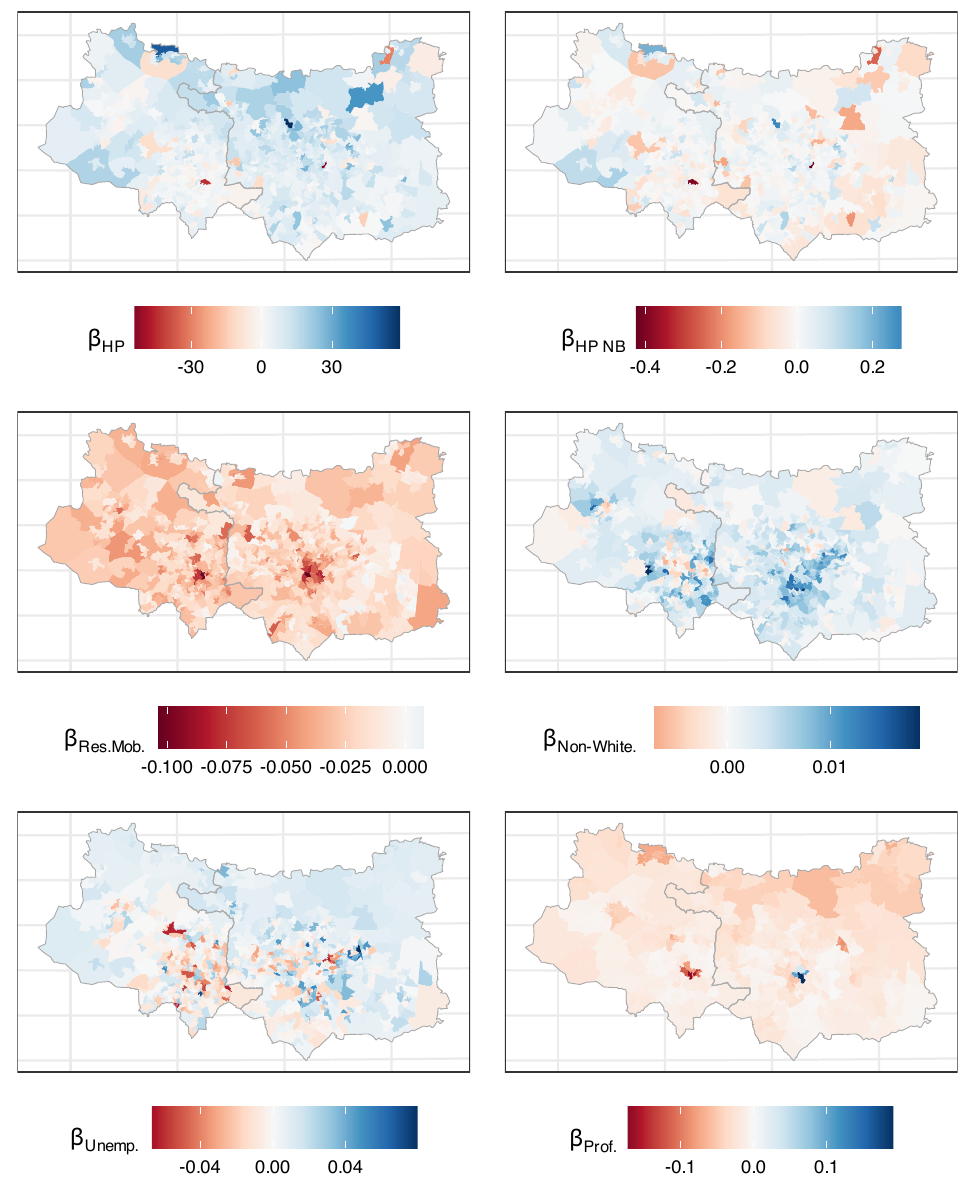} 

}

\caption{The spatial distribution of the change rates ($\beta_t$) across the Leeds and Bradford case study area.}\label{fig:beta_map}
\end{figure}

In summary, a series of changes in socio-economic variables over 6 years
were calculated using LSOAs as neighbourhood areas. This was done
through annual measures of 5 demographic variables and spatially located
house sales transactions data. From the latter 2 measures were
generated: changes in annual annual median house price (price per square
metre) and changes relative to the nearest 250 `nearby' LSOAs, in which
annual ratios of \emph{LSOA to nearby LSOA} were calculated. A
regression approach was used to generates rates of change (\(\beta\) s)
of each variable over time. The results are shown in Figure
\ref{fig:beta_map} and summarised in Table \ref{tab:var_corr}. As the
coefficient estimates for the relative house price change are highly
correlated with change in annual house price, and but less correlated
with the socio-economic variables, a decision was made to retain the
relative house price change and to drop simple house price change.
Recall that the aim is to unpick the different dimensions of
neighbourhood change which may be better captured by variables with
lower correlations. Finally, each of the variables was rescaled
(\(N(0,1)\)).

\begin{longtable}[]{@{}
  >{\raggedright\arraybackslash}p{(\linewidth - 12\tabcolsep) * \real{0.1591}}
  >{\raggedleft\arraybackslash}p{(\linewidth - 12\tabcolsep) * \real{0.1364}}
  >{\raggedleft\arraybackslash}p{(\linewidth - 12\tabcolsep) * \real{0.1364}}
  >{\raggedleft\arraybackslash}p{(\linewidth - 12\tabcolsep) * \real{0.1591}}
  >{\raggedleft\arraybackslash}p{(\linewidth - 12\tabcolsep) * \real{0.1136}}
  >{\raggedleft\arraybackslash}p{(\linewidth - 12\tabcolsep) * \real{0.1477}}
  >{\raggedleft\arraybackslash}p{(\linewidth - 12\tabcolsep) * \real{0.1477}}@{}}
\caption{\label{tab:var_corr}The corrleations between the variaous
changes in socio-economic variables.}\tabularnewline
\toprule\noalign{}
\begin{minipage}[b]{\linewidth}\raggedright
\end{minipage} & \begin{minipage}[b]{\linewidth}\raggedleft
House Price
\end{minipage} & \begin{minipage}[b]{\linewidth}\raggedleft
Relative HP
\end{minipage} & \begin{minipage}[b]{\linewidth}\raggedleft
Res. Mobility
\end{minipage} & \begin{minipage}[b]{\linewidth}\raggedleft
Non-White
\end{minipage} & \begin{minipage}[b]{\linewidth}\raggedleft
Unemployment
\end{minipage} & \begin{minipage}[b]{\linewidth}\raggedleft
Prof.~Occup.
\end{minipage} \\
\midrule\noalign{}
\endfirsthead
\toprule\noalign{}
\begin{minipage}[b]{\linewidth}\raggedright
\end{minipage} & \begin{minipage}[b]{\linewidth}\raggedleft
House Price
\end{minipage} & \begin{minipage}[b]{\linewidth}\raggedleft
Relative HP
\end{minipage} & \begin{minipage}[b]{\linewidth}\raggedleft
Res. Mobility
\end{minipage} & \begin{minipage}[b]{\linewidth}\raggedleft
Non-White
\end{minipage} & \begin{minipage}[b]{\linewidth}\raggedleft
Unemployment
\end{minipage} & \begin{minipage}[b]{\linewidth}\raggedleft
Prof.~Occup.
\end{minipage} \\
\midrule\noalign{}
\endhead
\bottomrule\noalign{}
\endlastfoot
House Price & 1.000 & 0.880 & 0.091 & -0.077 & 0.127 & -0.097 \\
Relative HP & 0.880 & 1.000 & -0.023 & 0.006 & 0.033 & -0.017 \\
Res. Mobility & 0.091 & -0.023 & 1.000 & -0.314 & 0.155 & 0.030 \\
Non-White & -0.077 & 0.006 & -0.314 & 1.000 & 0.012 & 0.115 \\
Unemployment & 0.127 & 0.033 & 0.155 & 0.012 & 1.000 & -0.101 \\
Prof.~Occup. & -0.097 & -0.017 & 0.030 & 0.115 & -0.101 & 1.000 \\
\end{longtable}

\newpage

\subsection{Social media data and
preprocessing}\label{social-media-data-and-preprocessing}

Social media data (SMD) for the Leeds and Bradford areas was analysed.
Approximately 5.1 million geolocated tweets for 2014 to 2019 were
extracted from the Twitter / X API for this area in West Yorkshire, UK
(i.e.~the period prior to the COVID-19 pandemic) before the API closed.
These data were not filtered or selected in any way (e.g.~for specific
terms or tags).

The Twitter dataset needed to be filtered for personal users because raw
social media streams can be overwhelmed by noise from automated
accounts, and, in this context, irrelevant content that may distort any
SMD analysis. Recall that the aim was to analyse the SMD in order to
extract geolocated sentiment, behaviour or trends and relate these over
neighbourhoods to house prices.

A large language model (LLM) was used to do this using the following
prompt:

\begin{Verbatim}[
  breaklines=true,
  breakanywhere=true,
  breaksymbolleft={},
  breaksymbolright={}
]

You are a classifier that decides whether tweets posted by an ORDINARY PERSON sharing personal experience, or by a COMMERCIAL or ORGANISATIONAL or AUTOMATED account.

I will provide you with a batch of 20 Twitter / X posts. Your task is to analyse the text in each Tweet and to classify the tweet into exactly one of these labels:

- "personal"    : an individual posting about their own life, opinions, feelings, social activity, observations of a place, complaints, humour, or interaction with friends.
- "commercial"  : promotion, advertising, marketing, sales, job adverts, property / estate-agent listings, event promotion by a venue or business, PR from an organisation, influencer sponsored content, or affiliate links.
- "news_media"  : posts from news outlets, journalists acting in a professional capacity, or official press releases.
- "automated"   : bot-generated content such as weather updates, traffic reports, air-quality readings, earthquake alerts, "I just ran X miles with Strava", Foursquare check-ins with no personal commentary, scheduled content, or repetitive template-driven posts.
- "public_org"  : councils, police, NHS, universities, charities, or other non-commercial institutions posting in an official capacity.
- "unclear"     : genuinely ambiguous — use sparingly.

Decision cues and rules (non-exhaustive):
* Commercial signals: prices, "for sale", "now hiring", "book now", "link in bio", "[pin emoji]venue", shop hours, menu items, hashtags like  #ad #sponsored #promo, URL to a storefront, property portals  (rightmove, zoopla, onthemarket), "our team", "our customers".
* Automated signals: exact templates repeated (e.g. "Wind: 5mph ..."), auto-generated check-ins with just a place + URL, app attribution strings ("via IFTTT", "Runkeeper", "Foursquare").
* Personal signals: first-person pronouns with subjective content, slang, emotional language, replies ("@someone lol same"), conversational fragments, personal photos described in text.

Return ONLY a single JSON object, no prose, no code fences:
* You must return exactly 20 labels, one per post, in a structured format.
* The output should be a JSON array of exactly 20 labels.
* If there are fewer or more than 20 labels, review and correct the response before 
returning it.
* Make sure the JSON array is correctly formatted before returning it.

\end{Verbatim}

This was passed to the meta-llama/Llama-3.3-70B-Instruct-Turbo model
(featuring FP8 quantisation) accessed via the Together AI serverless
inference API (AI 2024) in April 2026.

\begin{figure}

{\centering \includegraphics{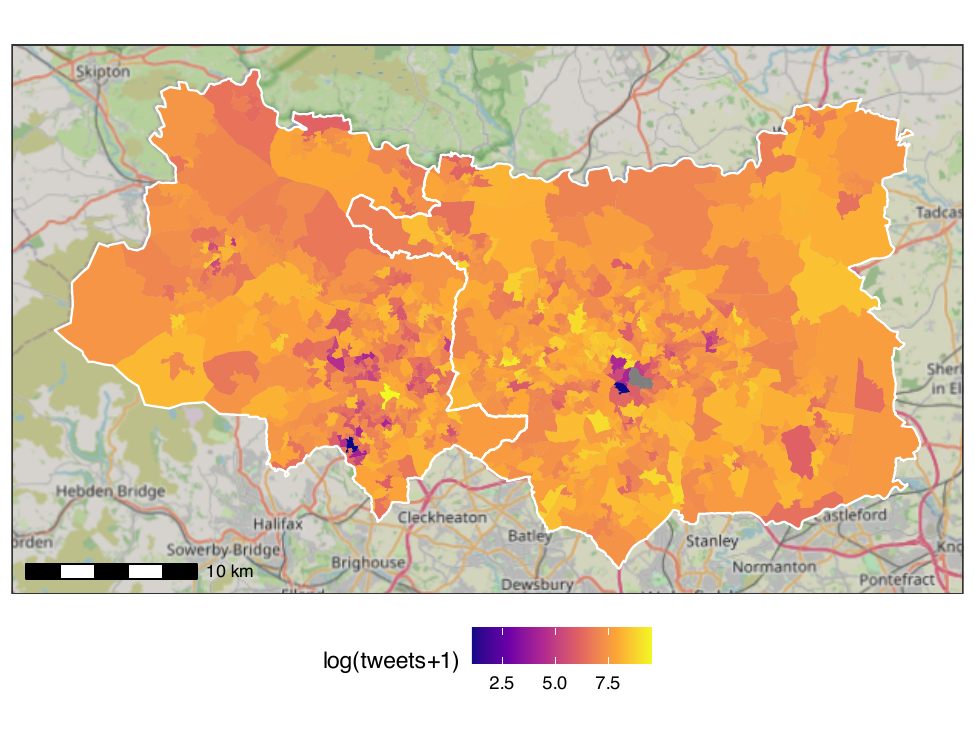} 

}

\caption{The spatial distribution of Tweet density per LSOA (2014–2019).}\label{fig:tweet-coverage-map}
\end{figure}

The filtered geolocated SMD were then linked to LSOA neighbourhood areas
through spatial intersection, with each tweet assigned to an LSOA to
support subsequent analyses at the LSOA level. The SMD filtering and
LSOA spatial intersection resulted in 1.6 million tweets, as shown in
Table \ref{tab:tweet_summary}. It is evident that the SMD distribution
is severely imbalanced, with approximately 72\% of all tweets from 2014,
with counts declining steeply through to 2019 (\textasciitilde1.2\% of
total). This is not a sampling artefact --- Twitter progressively
reduced geo-tagging functionality from 2015 onwards, and geolocated
tweet volumes dropped industry-wide. This temporal imbalance is not
addressed by filtering for personal users. This potentially indicates
the need to account for this imbalance in any analysis or model
construction. The spatial distribution of pre-processed SMD is shown in
Figure \ref{fig:tweet-coverage-map}.

\begin{longtable}[]{@{}lrrrrr@{}}
\caption{\label{tab:tweet_summary}Tweet counts per LSOA × year for Leeds
and Bradford.}\tabularnewline
\toprule\noalign{}
Year & LSOAs with tweets & Total tweets & Median/LSOA & Q25 & Q75 \\
\midrule\noalign{}
\endfirsthead
\toprule\noalign{}
Year & LSOAs with tweets & Total tweets & Median/LSOA & Q25 & Q75 \\
\midrule\noalign{}
\endhead
\bottomrule\noalign{}
\endlastfoot
2014 & 788 & 1,139,059 & 1,215 & 708.8 & 1,973.5 \\
2015 & 786 & 327,979 & 264 & 134.0 & 519.5 \\
2016 & 739 & 42,438 & 17 & 6.0 & 47.5 \\
2017 & 693 & 28,688 & 11 & 3.0 & 33.0 \\
2018 & 633 & 21,568 & 9 & 3.0 & 28.0 \\
2019 & 604 & 18,830 & 6 & 2.0 & 20.0 \\
\end{longtable}

In summary, filtering and spatially intersecting unrestricted geolocated
SMD to support subsequent neighbourhood (LSOA) level analyses resulted
in a final dataset of approximately 1.6 million geolocated tweets.
However, the data are highly temporally imbalanced, with a steep decline
from 2014 to 2019. This suggests that any annual SMD signal is
unreliable in later years for many LSOAs. There are two practical
responses to this: (i) aggregate over all years together to maximise
power; and / or (ii) apply a minimum tweet-count threshold per LSOA-year
before including an observation in analytical models.

\section{SMD Analysis}\label{smd-analysis}

A series of analyses of the SMD were undertaken: STM to identify latent
topics in the SMD, LLM scoring of neighbourhood characterisations and
text embeddings were to estimate semantic distances to gentrification
archetypes. The aim of each was to generate attributes from the SMD that
could be linked to the LSOA demographic and house price change
attributes. The three analyses described in this section generated LSOA
scores of some kind: 9 topics from the STM analysis, 5 scores from the
LLM neighbourhood characterisation and 3 scores from the text embeddings
analysis. These were combined into a single spatial dataset, with each
record representing an LSOA. This was to be passed to the analyses in
the in Sections 4.

\subsection{Structural Topic Models
(STM)}\label{structural-topic-models-stm}

Structural Topic Models (Roberts et al. 2015; Blei et al. 2003) were
applied to the SMD. They are an extension of Latent Dirichlet Allocation
(Blei et al. 2003) and allow topic prevalence to vary as a function of
document-level metadata. In the STM framework, each document \(d\)
(here, an LSOA × year combination of aggregated social media data) was
modelled as a mixture of \(K\) topics, generated through a
logistic-normal model in which topic proportions are drawn on the
log-odds scale and then mapped onto the topic simplex:

\begin{equation}
\eta_d \sim \mathcal{N}(\mu_d, \Sigma), \qquad \theta_d = g(\eta_d)
\label{stm1}
\end{equation}

where \(\eta_d = (\eta_{d,1}, \ldots, \eta_{d,K})\) is a
document-specific vector on the log-odds scale, \(\Sigma\) is a
\(K \times K\) covariance matrix that allows topics to correlate with
one another, \(g\) is the softmax (logistic-normal) transformation
mapping \(\eta_d\) onto the \(K\)-simplex, and
\(\theta_d = (\theta_{d,1}, \ldots, \theta_{d,K})\) is the resulting
vector of topic proportions for document \(d\). Unlike standard LDA,
where the mean of this distribution is fixed across documents, the
prevalence covariate specification allows the mean \(\mu_d\) to vary
with document-level metadata:

\begin{equation}
\mu_d = \gamma + X_d \beta
\label{stm2}
\end{equation}

where \(X_d\) is the vector of covariates for document \(d\), \(\beta\)
is the coefficient vector, and \(\gamma\) is an intercept. In this
analysis, the prevalence model was specified as:

\begin{equation}
\mu_d = \gamma + \beta_{hp} X_{hp,d} + \beta_{rm} X_{rm,d} + \beta_{nw} X_{nw,d} + \beta_{un} X_{un,d} + \beta_{pr} X_{pr,d} + s(year_d)
\label{stm3}
\end{equation}

where \(X_{hp,d}\), \(X_{rm,d}\), \(X_{nw,d}\), \(X_{un,d}\) and
\(X_{pr,d}\) are, respectively, the standardised relative house price
change, change in residential mobility, change in the non-white resident
population, change in unemployment, and change in professional
occupation for the LSOA associated with document \(d\). \(s(year_d)\) is
a penalised thin-plate regression spline allowing non-linear time
effects (Roberts et al. 2015). The terms \(\beta_{hp}\), \(\beta_{rm}\),
\(\beta_{nw}\), \(\beta_{un}\) and \(\beta_{pr}\) are the coefficients
of interest, capturing how topic proportions vary with neighbourhood
house price dynamics and with neighbourhood change in residential
mobility, ethnic composition, unemployment and professional occupation.
By including these terms as prevalence covariates, the STM directly
estimates which topics are more or less prevalent in LSOAs with high or
low values of each change measure, without requiring the topic
vocabulary to be pre-specified. Topics are discovered from the data
using spectral initialisation (Arora et al. 2012) and can be interpreted
post-hoc with respect to each of the five \(\beta\) coefficients.

The Topic Prevalence model (Equation \ref{stm3}) yields coefficient
estimates \(\beta_{k,j}\) for each topic \(k\) and each of the five
socio-economic predictors \(j\), quantifying how topic proportions vary
with each dimension of neighbourhood change. Formally, for topic \(k\),
the regression coefficient \(\beta_{k,j}\) in the logistic-normal model
is interpreted as:

\begin{equation}
\frac{\partial \mu_{d,k}}{\partial X_{j,d}} = \beta_{k,j}
\label{stm4}
\end{equation}

where \(X_{j,d}\) is the standardised value of predictor \(j\) for
document \(d\): \(hp\) (relative house price change), \(rm\) (change in
residential mobility), \(nw\) (change in the non-white resident
population), \(un\) (change in unemployment), and \(pr\) (change in
professional occupation). Under the convention that each \(X_{j,d}\) is
standardised (zero mean, unit variance), \(\beta_{k,j}\) can be
interpreted as the expected change in log-odds of topic \(k\) per
standard-deviation increase in predictor \(j\), holding the other four
predictors constant.

Under this specification:

\begin{itemize}
\tightlist
\item
  A positive coefficient \(\beta_{k,j} > 0\) indicates that topic \(k\)
  is more prevalent in LSOAs where predictor \(j\) increased faster than
  the LSOA-wide average over the study period.
\item
  A negative coefficient \(\beta_{k,j} < 0\) indicates that topic \(k\)
  is more prevalent in LSOAs where predictor \(j\) increased more
  slowly, stagnated, or declined relative to other LSOAs.
\end{itemize}

For \(hp\_beta\) specifically, this retains the original reading: LSOAs
with \(\beta_{k,hp} > 0\) sit in the gentrifying tail of relative
house-price change (\(X_{hp,d} > \bar{X}_{hp}\)), while
\(\beta_{k,hp} < 0\) sit in the declining tail
(\(X_{hp,d} < \bar{X}_{hp}\)). The same directional logic extends to the
other four predictors, each capturing a distinct facet of neighbourhood
socio-economic change: rising residential mobility, demographic
turnover, labour-market distress, and professionalisation of the
resident population.

This directly tests the question: does social media discourse
systematically differ across LSOAs undergoing different trajectories of
house-price change, residential mobility, demographic turnover,
unemployment, and occupational change? Each predictor's sign and
magnitude, for a given topic, tell us which type of neighbourhood change
is more likely to be associated with discourse on that theme. For each
of the K=20 topics and each of the 5 predictors, we test the null
hypothesis:

\begin{equation}
H_{0,k,j}: \beta_{k,j} = 0 \quad \text{(topic k prevalence does not vary with predictor } j\text{)}
\label{stm5}
\end{equation}

versus the two-sided alternative \(H_{1,k,j}: \beta_{k,j} \neq 0\). The
test statistic is the t-ratio
\(t_{k,j} = \hat{\beta}_{k,j} / \hat{SE}(\hat{\beta}_{k,j})\), and
p-values were computed under a Student's t-distribution with degrees of
freedom determined by the effective sample size accounting for the
estimation uncertainty in the STM parameters (Roberts et al. 2015;
Bingham and Fry 2013).

Because \(5 \times 20 = 100\) hypotheses are tested across the five
predictors, there is a risk of false discoveries due to multiple
comparisons. Following Benjamini and Hochberg (1995), the False
Discovery Rate (FDR) is controlled at 5\% separately within each
predictor's family of 20 topic-level tests, rather than pooling all 100
tests into a single correction, since each predictor addresses a
substantively distinct question about which type of neighbourhood change
is associated with topic prevalence. For each predictor \(j\), the
critical p-value threshold is computed as:

\begin{equation}
p^*_j = \frac{\alpha \cdot r}{K}
\label{stm6}
\end{equation}

where \(\alpha = 0.05\) is the target FDR, \(r\) is the rank of the
p-value when sorted in ascending order within predictor \(j\)'s 20
tests, and K=20 is the number of topics. This ensures that, for each
predictor, in expectation no more than 5\% of topics declared
significant for that predictor are false discoveries.

Using this correction, for a given predictor \(j\), a topic is deemed to
be significantly associated with predictor \(j\) at \(FDR \le 5\%\) if
its Benjamini--Hochberg-adjusted p-value (q-value) is
\(\le \alpha = 0.05\), computed within that predictor's family of
\(K=20\) topic-level tests. This procedure is applied independently to
each of the five predictors, and this conservative approach accounts for
the fact that many hypotheses are being tested from a single model.

Table \ref{tab:stm_topic_assoc} shows the significant associations
between STM topics and the socio-economic rates of change. It includes
the Benjamini--Hochberg-adjusted p-value (the q-value column) with
significance determined by whether this falls at or below the target FDR
of 5\%, within that predictor's family of \(K=20\) topic-level tests.
Because the correction is applied separately to each predictor rather
than pooled across all five (see above), significance is
predictor-specific --- a topic may be flagged for one predictor and not
another. Thus the table should be read predictor-by-predictor rather
than down the table as a whole. A further summary of the significant
topics grouped by the socio-economic rates of change is shown in Table
\ref{tab:stm_topic_assoc_sum}.

\begin{longtable}[]{@{}
  >{\raggedright\arraybackslash}p{(\linewidth - 12\tabcolsep) * \real{0.2469}}
  >{\raggedleft\arraybackslash}p{(\linewidth - 12\tabcolsep) * \real{0.0741}}
  >{\raggedright\arraybackslash}p{(\linewidth - 12\tabcolsep) * \real{0.1111}}
  >{\raggedright\arraybackslash}p{(\linewidth - 12\tabcolsep) * \real{0.1358}}
  >{\raggedright\arraybackslash}p{(\linewidth - 12\tabcolsep) * \real{0.2346}}
  >{\raggedright\arraybackslash}p{(\linewidth - 12\tabcolsep) * \real{0.0988}}
  >{\raggedright\arraybackslash}p{(\linewidth - 12\tabcolsep) * \real{0.0988}}@{}}
\caption{\label{tab:stm_topic_assoc} The significant associations
between STM topics and the different socio-economic rates of
change.}\tabularnewline
\toprule\noalign{}
\begin{minipage}[b]{\linewidth}\raggedright
Predictor
\end{minipage} & \begin{minipage}[b]{\linewidth}\raggedleft
Topic
\end{minipage} & \begin{minipage}[b]{\linewidth}\raggedright
Estimate
\end{minipage} & \begin{minipage}[b]{\linewidth}\raggedright
Std. Error
\end{minipage} & \begin{minipage}[b]{\linewidth}\raggedright
95\% CI
\end{minipage} & \begin{minipage}[b]{\linewidth}\raggedright
p-value
\end{minipage} & \begin{minipage}[b]{\linewidth}\raggedright
q-value
\end{minipage} \\
\midrule\noalign{}
\endfirsthead
\toprule\noalign{}
\begin{minipage}[b]{\linewidth}\raggedright
Predictor
\end{minipage} & \begin{minipage}[b]{\linewidth}\raggedleft
Topic
\end{minipage} & \begin{minipage}[b]{\linewidth}\raggedright
Estimate
\end{minipage} & \begin{minipage}[b]{\linewidth}\raggedright
Std. Error
\end{minipage} & \begin{minipage}[b]{\linewidth}\raggedright
95\% CI
\end{minipage} & \begin{minipage}[b]{\linewidth}\raggedright
p-value
\end{minipage} & \begin{minipage}[b]{\linewidth}\raggedright
q-value
\end{minipage} \\
\midrule\noalign{}
\endhead
\bottomrule\noalign{}
\endlastfoot
\(HousePrice_\beta\) & 4 & -0.0081 & 0.0025 & {[}-0.0129, -0.0032{]} &
0.0011 & 0.0222 \\
\(NonWhite_\beta\) & 12 & 0.0102 & 0.0025 & {[}0.0053, 0.0151{]} &
0.0000 & 0.0010 \\
\(NonWhite_\beta\) & 14 & -0.0085 & 0.0023 & {[}-0.0130, -0.0040{]} &
0.0002 & 0.0020 \\
\(NonWhite_\beta\) & 5 & 0.0121 & 0.0036 & {[}0.0050, 0.0193{]} & 0.0009
& 0.0059 \\
\(NonWhite_\beta\) & 18 & 0.0096 & 0.0035 & {[}0.0028, 0.0164{]} &
0.0058 & 0.0291 \\
\(Prof.Occup._\beta\) & 7 & -0.0134 & 0.0032 & {[}-0.0197, -0.0071{]} &
0.0000 & 0.0006 \\
\(Prof.Occup._\beta\) & 18 & 0.0125 & 0.0032 & {[}0.0062, 0.0188{]} &
0.0001 & 0.0009 \\
\(Prof.Occup._\beta\) & 16 & 0.0124 & 0.0037 & {[}0.0051, 0.0197{]} &
0.0009 & 0.0059 \\
\(Prof.Occup._\beta\) & 6 & -0.0106 & 0.0034 & {[}-0.0173, -0.0040{]} &
0.0017 & 0.0085 \\
\(Prof.Occup._\beta\) & 11 & -0.0107 & 0.0039 & {[}-0.0183, -0.0032{]} &
0.0055 & 0.0219 \\
\(Res.Mob._\beta\) & 7 & -0.0192 & 0.0028 & {[}-0.0247, -0.0136{]} &
0.0000 & 0.0000 \\
\(Res.Mob._\beta\) & 16 & -0.0166 & 0.0041 & {[}-0.0245, -0.0086{]} &
0.0000 & 0.0005 \\
\(Res.Mob._\beta\) & 6 & 0.0094 & 0.0033 & {[}0.0029, 0.0158{]} & 0.0045
& 0.0301 \\
\(Res.Mob._\beta\) & 17 & -0.0035 & 0.0014 & {[}-0.0062, -0.0007{]} &
0.0127 & 0.0499 \\
\(Res.Mob._\beta\) & 12 & 0.0062 & 0.0025 & {[}0.0013, 0.0111{]} &
0.0134 & 0.0499 \\
\(Res.Mob._\beta\) & 5 & 0.0087 & 0.0036 & {[}0.0017, 0.0158{]} & 0.0150
& 0.0499 \\
\(Unemployed_\beta\) & 7 & -0.0085 & 0.0026 & {[}-0.0135, -0.0034{]} &
0.0011 & 0.0219 \\
\end{longtable}

\begin{longtable}[]{@{}ll@{}}
\caption{\label{tab:stm_topic_assoc_sum} A summary of the Topics
significantly associated with each socio-economic rate of
change.}\tabularnewline
\toprule\noalign{}
Predictor & Topics \\
\midrule\noalign{}
\endfirsthead
\toprule\noalign{}
Predictor & Topics \\
\midrule\noalign{}
\endhead
\bottomrule\noalign{}
\endlastfoot
\(HousePrice_\beta\) & 4 \\
\(NonWhite_\beta\) & 5, 12, 14, 18 \\
\(Prof.Occup._\beta\) & 6, 7, 11, 16, 18 \\
\(Res.Mob._\beta\) & 5, 6, 7, 12, 16, 17 \\
\(Unemployed_\beta\) & 7 \\
\end{longtable}

As a final evaluation, the top 100 words for 10 of the topics that were
significantly associated with the different socio-economic changes were
passed to an LLM (Anthropic 2026) for an interpretation as detailed in
Supplementary materials (Table \ref{tab:topic_themes}). The results are
summarised in Table \ref{tab:stm_summary} and include the LLM generated
labels, key themes, caveats and links to socio-economic changes.

\begin{itemize}
\tightlist
\item
  Topic 4: UK Political \& Religious Debate. Associated with house price
  change, suggesting that political, religious and international debate
  is more prominent in areas experiencing differing trajectories in
  relative house prices.
\item
  Topic 5: Casual Youth Social Chatter. Associated with change in the
  non-white population, although its generic nature makes this more
  likely to reflect demographic or social-media composition than a
  specific neighbourhood process.
\item
  Topic 6: Gratitude \& Well-Wishing Chatter. Associated with
  professional occupation and residential mobility, indicating a
  relationship with changing professionalisation and population
  turnover, although the generic content makes interpretation uncertain.
\item
  Topic 7: Bradford British-Pakistani Community Life. Associated with
  professional occupation changes, residential mobility and
  unemployment, linking discussion of an established British-Pakistani
  community to areas characterised by relatively low or changing
  professionalisation, mobility and employment.
\item
  Topic 11: Rural Yorkshire Leisure \& Heritage. Associated with
  professional occupation changes, with countryside, heritage and
  leisure discourse occurring alongside changes in the professional
  occupational profile of neighbourhoods.
\item
  Topic 12: Leeds United Football Fandom. Associated with change in the
  non-white population, suggesting that football-related discourse is
  concentrated in neighbourhoods experiencing particular demographic
  trajectories, rather than implying a substantive relationship with
  football itself.
\item
  Topic 14: Pennine Village Heritage \& Food. Associated with changes in
  the non-white population, linking heritage, food and local-identity
  discourse with changing ethnic composition in relatively low-diversity
  Pennine communities.
\item
  Topic 16: Youth \& Student Social Chat. Associated with changes in
  professional occupation and residential mobility, suggesting that
  informal youth/student discourse occurs in neighbourhoods undergoing
  professionalisation alongside relatively low or changing residential
  turnover.
\item
  Topic 17: Christian Worship \& Prayer. Associated with residential
  mobility, suggesting that this bilingual English/French devotional and
  worship discourse is more prominent in lower-turnover, more
  residentially stable neighbourhoods.
\item
  Topic 18: General Personal \& Family Chat. Associated with non-white
  population and professional occupation changes, although its generic
  content provides little evidence for a specific neighbourhood-change
  mechanism.
\end{itemize}

Of these 10 topics, 9 were selected for inclusion in later statistical
analysis. Topic 17 was dropped due to its extremely niche nature.

It is possible to cautiously construct a narrative around each topic
based on the relationships between topic prevalence and rates of
socio-economic change. Considered together, the statistical associations
and the characterisation of the topics suggest that different dimensions
of neighbourhood change may be associated with different social-media
discourse, rather than a single social-media trace of neighbourhood
change being present. Some associations are geographically and
culturally distinctive, particularly Topic 7, whereas others involve
generic social-media activity and provide weaker evidence. This supports
the approach whereby social-media topics are treated as multiple,
partial indicators of neighbourhood dynamics, rather than direct
measures of processes such as gentrification, decline etc.

\begin{landscape}
\begin{longtable}[t]{>{\raggedleft\arraybackslash}p{0.7cm}>{\raggedright\arraybackslash}p{3cm}>{\raggedright\arraybackslash}p{5cm}>{\raggedright\arraybackslash}p{8cm}>{\raggedright\arraybackslash}p{2.5cm}}
\caption{\label{tab:unnamed-chunk-4}\label{tab:stm_summary}LLM generated interpretations of the STM topics that were significantly associated with socio-economic changes, with a summary label, key themes, caveats and the socio-economic changes they are associated with.}\\
\toprule
Topic & Label & Key Themes & Caveats & Socio-Economic Changes\\
\midrule
4 & UK Political \& Religious Debate & UK party politics, broadcast news commentary, and contested international and religious affairs discussed in an opinionated register & Topic conflates domestic politics with religion and foreign affairs; the house-price association may reflect general political engagement in less-affluent areas rather than gentrification-specific discourse, and causality cannot be inferred from cross-sectional STM coefficients. & House Price\\
5 & Casual Youth Social Chatter & Informal, expletive-heavy everyday chatter with no fixed subject, typical of younger social-media users & Very low thematic specificity — this is largely generic platform chatter; the predictor significance may reflect population age structure and volume rather than a substantive discourse pattern. & Non White, Res. Mob.\\
6 & Gratitude \& Well-Wishing Chatter & Polite, low-content social courtesy and reciprocal well-wishing, not tied to a specific subject & Highly generic phatic content shared across many social-media topics; weak thematic specificity limits substantive interpretation, and the mixed-sign associations across predictors make one coherent neighbourhood story hard to justify. & Prof. Occup., Res. Mob.\\
7 & Bradford British-Pakistani Community Life & Bradford's British-Pakistani community life, food culture, and religious observance rooted in a stable, low-turnover population & Negative coefficients indicate below-average change, not decline per se; an already-settled community with low baseline mobility could show this pattern without any deterioration, so 'stagnation' should not be read as decline. & Prof. Occup., Res. Mob., Unemployed\\
11 & Rural Yorkshire Leisure \& Heritage & Countryside leisure, heritage villages, and outdoor lifestyle content in already-affluent, demographically stable locations & A negative coefficient here may reflect a ceiling effect — areas already high in non-white share or professional occupation cannot rise much further — rather than genuine decline; avoid reading sign as deprivation. & Prof. Occup.\\
\addlinespace
12 & Leeds United Football Fandom & Leeds United match-day commentary and fan culture, concentrated in neighbourhoods with faster demographic turnover & Association is with turnover and demographic composition, not football itself; correlation may reflect the age/renter profile of these LSOAs rather than any substantive link between fandom and neighbourhood change. & Non White, Res. Mob.\\
14 & Pennine Village Heritage \& Food & Brontë-country heritage tourism, home cooking, and local identity in historically low-diversity Pennine villages & Single-predictor significance (nw\_beta only) limits how much can be claimed; the association describes the relative rate of change in ethnic composition, not the area's current diversity level. & Non White\\
16 & Youth \& Student Social Chat & Informal student and young-adult chatter concentrated in settled, professionalising neighbourhoods with low turnover & Generic conversational content with weak topical specificity; the low-mobility/high-professionalisation combination is unusual and warrants checking against ward-level context rather than taking at face value. & Prof. Occup., Res. Mob.\\
17 & Christian Worship \& Prayer & Christian worship, prayer, and praise discourse spanning English and French, concentrated in stable, lower-turnover faith communities & Significant on residential mobility only; the unusual English-French language mix and the isolated 'keighley' reference warrant checking the underlying documents before attributing this to any specific congregation or community. & Res. Mob.\\
18 & General Personal \& Family Chat & Generic everyday personal and family chatter, weakly linked to several, partly conflicting change measures & Significant on three predictors simultaneously with no clear substantive theme; likely reflects general discourse volume/density rather than a specific socio-economic mechanism — interpret with caution. & Non White, Prof. Occup.\\
\bottomrule
\end{longtable}
\end{landscape}

\subsection{LLM-Based Neighbourhood
Characterisation}\label{llm-based-neighbourhood-characterisation}

The social media data were also passed directly to an LLM (Anthropic
2026) with a query to characterise the neighbourhood over five
theoretically motivated dimensions. The tweets were aggregated to the
LSOA × year level rather than scoring individual tweets. This is
potentially more sensitive than scoring individual tweets because the
LLM can detect patterns across many posts and produce features directly
linked to theoretical constructs in the neighbourhood change literature.
The prompt passed to the LLM is below. It asked the LLM to score the
aggregated SMD in 5 dimensions, with the objective of extracting
meaningful measures that could be linked statistically to LSOA level HP
changes (\(\beta_t\)).

\begin{Verbatim}[
  breaklines=true,
  breakanywhere=true,
  breaksymbolleft={},
  breaksymbolright={}
]

You are an urban analyst examining social media posts to characterise the neighbourhood from which they were posted.

I will provide a sample of tweets from a small residential area (UK Lower Super Output Area) in a specific year. Assess the neighbourhood on FIVE dimensions, each scored 1-10.

DIMENSIONS:
1. ECONOMIC_VITALITY:      Signs of new businesses, investment, economic activity, employment. 1 = stagnant/declining; 10 = growing/vibrant.
2. DISPLACEMENT_PRESSURE:  Affordability stress, rent rises, evictions, people being priced out. 1 = none evident; 10 = severe.
3. SOCIAL_COHESION:        Community bonds, local events, shared neighbourhood identity. 1 = fragmented; 10 = very strong.
4. PHYSICAL_ENVIRONMENT:   Built environment quality, streets, green space, cleanliness. 1 = very poor; 10 = very good.
5. DEMOGRAPHIC_CHANGE:     Changing population composition, new resident types, cultural/linguistic shift. 1 = none evident; 10 = very marked.

Return ONLY valid JSON — no other text — in exactly this format:
{
  \"economic_vitality\": X,
  \"displacement_pressure\": X,
  \"social_cohesion\": X,
  \"physical_environment\": X,
  \"demographic_change\": X,
  \"confidence\": X
}

The 'confidence' field (1-10) reflects how much relevant neighbourhood information the tweets contain. If tweets provide no useful signal, return 5 for all dimensions and 1 for confidence.

Return ONLY the raw JSON object with 6 elements. Do not wrap it in markdown code fences.
Do not include the word 'json' or any backticks. 

Your entire response must start with { and end with }.

\end{Verbatim}

The resulting LSOA by year by dimension data were aggregated over each
LSOA to construct a weighted average for each of the 5 dimensions, with
each year's value weighted proportionally to its tweet count, giving
more influence to years with more observations. The correlations across
the dimensions and with socio-economic \(\beta_t\)s are shown in Table
\ref{tab:llm_corr}. Some of the dimensions show weak correlations with
the demographic and house price changes, but overall the correlations
between these and the LLM derived scores are weak, except for those with
changes in Unemployed. The LSOA values were retained for later
statistical analysis.

\begingroup\fontsize{9}{11}\selectfont

\begin{landscape}
\begin{longtable}[t]{lllllllllll}
\caption{\label{tab:llm-aggregate}\label{tab:llm_corr}Spearman correlations of LLM neighbourhood dimensions with changes in socio-economic variables.}\\
\toprule
 & House Price & Res. Mob. & Non White & Unemployed & Prof. Occup. & Econ. vitality & Displacement & Soc. Cohesion & Phys. Env. & Demographic\\
\midrule
House Price &  & -0.288 & 0.186 & -0.015 & 0.010 & -0.082 & -0.010 & -0.004 & -0.036 & -0.032\\
Res. Mob. &  &  & -0.035 & 0.213 & 0.028 & -0.051 & 0.007 & -0.048 & -0.059 & 0.004\\
Non White &  &  &  & -0.175 & 0.039 & -0.010 & -0.015 & 0.033 & 0.018 & -0.023\\
Unemployed &  &  &  &  & -0.016 & -0.127 & 0.223 & -0.266 & -0.277 & 0.206\\
Prof. Occup. &  &  &  &  &  & 0.040 & -0.005 & 0.007 & 0.036 & 0.030\\
\addlinespace
Econ. vitality &  &  &  &  &  &  & 0.370 & -0.020 & 0.483 & 0.464\\
Displacement &  &  &  &  &  &  &  & -0.769 & -0.203 & 0.927\\
Soc. Cohesion &  &  &  &  &  &  &  &  & 0.471 & -0.700\\
Phys. Env. &  &  &  &  &  &  &  &  &  & -0.149\\
Demographic &  &  &  &  &  &  &  &  &  & \\
\bottomrule
\end{longtable}
\end{landscape}
\endgroup{}

\subsection{Text Embeddings}\label{text-embeddings}

Embeddings were used to map each tweet into a high-dimensional semantic
space; cosine similarity to archetype texts representing gentrification
and decline provides a data-driven neighbourhood score that captures
synonyms and paraphrases missed by lexicons. This approach requires no
external API and runs entirely locally. The model is trained on the full
tweet corpus, so the embedding space reflects the vocabulary and
semantic relationships in this specific dataset.

This analysis implemented a semantic embedding-based document
classification system that scores tweets along a gentrification-decline
continuum using word embeddings and centroid-based concept matching. It
applied a skip-gram word-to-vector model (Mikolov, Chen, et al. 2013)
that maps words from vocabulary \(V\) to \(\mathbb{R}^{100}\). The
training objective seeks to maximise:

\begin{equation}
\frac{1}{T} \sum_{t=1}^{T} \sum_{-c \leq j \leq c, j \neq 0} \log P(w_{t+j} | w_t; \theta)
\label{embed1}
\end{equation}

with context window \(c = 5\). Negative sampling (Mikolov, Sutskever, et
al. 2013) with \(k = 5\) samples reduces computational complexity. Words
appearing fewer than 5 times are excluded.

Seed word selection and vocabulary filtering was applied. This defined
two conceptual categories of gentrification and decline using specified
seed word sets, following the lexicon-based approach to sentiment and
topic analysis (Liu 2012; Taboada et al. 2011). The gentrification seed
words include terms such as ``gentrification'', ``investment'',
``revitalise'', ``regeneration'', ``redevelopment'', ``boutique'',
``artisan'', ``upmarket'', ``renovation'', ``improvement'', ``trendy'',
``affluent'', ``luxury'', and ``hipster.'' The decline seed words
include ``decline'', ``derelict'', ``abandoned'', ``poverty'',
``crime'', ``eviction'', ``displacement'', ``deprived'',
``unemployment'', ``rundown'', ``neglect'', ``vandalism'',
``antisocial'', and ``deprivation''.

Importantly, both seed word sets were intersected with the learned
vocabulary. In this case, two conceptual categories were defined as
intersections with the learned vocabulary:

\begin{equation}
S_{\text{gent}} = \{\text{gentrification, investment, revitalise, ...}\} \cap V
\label{embed2}
\end{equation}

\begin{equation}
S_{\text{decl}} = \{\text{decline, derelict, abandoned, ...}\} \cap V
\label{embed3}
\end{equation}

Having established the seed word vocabularies, two archetype centroid
vectors were constructed to represent the prototype meaning of each
concept. This follows the centroid-based approach to word sense
disambiguation and concept representation (Mitchell and Lapata 2010).
The gentrification centroid is computed as the mean of all embedding
vectors corresponding to gentrification seed words, with prototype
vectors computed as mean embeddings:

\begin{equation}
\mathbf{c}_{\text{gent}} = \frac{1}{|S_{\text{gent}}|} \sum_{w \in S_{\text{gent}}} \mathcal{M}(w)
\label{embed4}
\end{equation}

Similarly, the decline centroid is computed as:

\begin{equation}
\mathbf{c}_{\text{decl}} = \frac{1}{|S_{\text{decl}}|} \sum_{w \in S_{\text{decl}}} \mathcal{M}(w)
\label{embed5}
\end{equation}

where \(\mathcal{M}(w)\in  \mathbb{R}^{100}\) denotes the embedding
vector for word \(w\). This averaging approach creates points in the
embedding space that capture the typical semantic features associated
with each concept. By construction, semantically similar words (those
positioned nearby in the embedding space) will contribute similar
directional information to these centroids. This method extends
classical work in vector space models (Salton et al. 1975) to modern
distributed representations.

Document-level embeddings were used to transform each tweet into a
single vector representation by averaging the embeddings of all words it
contains, a technique known as the ``paragraph vector'' or doc2vec
average pooling approach (Le and Mikolov 2014; Arora et al. 2017). For a
tweet \(i\) containing a set of words \(T_i\), the tweet embedding is
computed as:

\begin{equation}
\mathbf{t}_i = \frac{1}{|T_i|} \sum_{w \in T_i} \mathcal{M}(w)
\label{embed6}
\end{equation}

This ``bag-of-words'' averaging approach produces a single
100-dimensional vector that summarises the semantic content of the
entire tweet. Arora et al. (2017) provide theoretical justification for
this averaging operation, showing that simple averaging of word
embeddings often captures the principal semantic components of a
document. This representation discards word order information but
captures the overall thematic composition of the document in a form
directly comparable to the archetype centroids.

To quantify the semantic relationship between each tweet and the two
archetypes, the analysis computes normalized dot-product similarity,
commonly known as cosine similarity (Salton and McGill 1983). The
similarity between two vectors \(\mathbf{a}\) and \(\mathbf{b}\) is
defined as:

\begin{equation}
\text{sim}(\mathbf{a}, \mathbf{b}) = \frac{\mathbf{a} \cdot \mathbf{b}}{|\mathbf{a}|_2 |\mathbf{b}|_2 + \epsilon}
\label{embed7}
\end{equation}

where \(|\cdot |_2\) denotes the L2 (Euclidean) norm and
\(\epsilon = 10^{-10}\) prevents numerical instability from division by
zero. Formally, the dot product is defined as:

\begin{equation}
\mathbf{a} \cdot \mathbf{b} = \sum_{k=1}^{d} a_k b_k
\label{embed8}
\end{equation}

and the L2 norm as:

\begin{equation}
|\mathbf{a}|_2 = \sqrt{\sum_{k=1}^{d} a_k^2}
\label{embed9}
\end{equation}

The cosine similarity metric ranges from -1 to 1, where 1 indicates
perfect alignment (vectors pointing in the same direction), 0 indicates
orthogonality (no correlation), and -1 indicates perfect opposition.
This measure is invariant to vector magnitude and depends only on the
angle between vectors, making it particularly suitable for comparing
high-dimensional embeddings (Turney and Pantel 2010).

Three complementary scores were generated for each tweet for
gentrification, decline and differential score. The gentrification and
decline scores measured the cosine similarity between each tweet's
embedding and the respective archetype, quantifying how closely its
semantic content was aligned with gentrification-related language. A
differential score was computed to indicate whether each tweet contained
language semantically closer to gentrification concepts (positive
scores) or to decline concepts (negative scores).

Finally, the embeddings scores were aggregated (median values) over each
LSOA and the correlations of these with HP \(\beta_t\) are shown in
Table \ref{tab:embed_corr}. The values for each LSOA were retained for
later statistical analysis.

\begin{longtable}[]{@{}
  >{\raggedright\arraybackslash}p{(\linewidth - 16\tabcolsep) * \real{0.1402}}
  >{\raggedleft\arraybackslash}p{(\linewidth - 16\tabcolsep) * \real{0.1121}}
  >{\raggedleft\arraybackslash}p{(\linewidth - 16\tabcolsep) * \real{0.0935}}
  >{\raggedleft\arraybackslash}p{(\linewidth - 16\tabcolsep) * \real{0.0935}}
  >{\raggedleft\arraybackslash}p{(\linewidth - 16\tabcolsep) * \real{0.1028}}
  >{\raggedleft\arraybackslash}p{(\linewidth - 16\tabcolsep) * \real{0.1215}}
  >{\raggedleft\arraybackslash}p{(\linewidth - 16\tabcolsep) * \real{0.1402}}
  >{\raggedleft\arraybackslash}p{(\linewidth - 16\tabcolsep) * \real{0.0748}}
  >{\raggedleft\arraybackslash}p{(\linewidth - 16\tabcolsep) * \real{0.1215}}@{}}
\caption{\label{tab:embed_corr}Spearman correlations of Embedding scores
with changes in socio-economic variables.}\tabularnewline
\toprule\noalign{}
\begin{minipage}[b]{\linewidth}\raggedright
\end{minipage} & \begin{minipage}[b]{\linewidth}\raggedleft
House Price
\end{minipage} & \begin{minipage}[b]{\linewidth}\raggedleft
Res. Mob.
\end{minipage} & \begin{minipage}[b]{\linewidth}\raggedleft
Non White
\end{minipage} & \begin{minipage}[b]{\linewidth}\raggedleft
Unemployed
\end{minipage} & \begin{minipage}[b]{\linewidth}\raggedleft
Prof.~Occup.
\end{minipage} & \begin{minipage}[b]{\linewidth}\raggedleft
Gentrification
\end{minipage} & \begin{minipage}[b]{\linewidth}\raggedleft
Decline
\end{minipage} & \begin{minipage}[b]{\linewidth}\raggedleft
Differential
\end{minipage} \\
\midrule\noalign{}
\endfirsthead
\toprule\noalign{}
\begin{minipage}[b]{\linewidth}\raggedright
\end{minipage} & \begin{minipage}[b]{\linewidth}\raggedleft
House Price
\end{minipage} & \begin{minipage}[b]{\linewidth}\raggedleft
Res. Mob.
\end{minipage} & \begin{minipage}[b]{\linewidth}\raggedleft
Non White
\end{minipage} & \begin{minipage}[b]{\linewidth}\raggedleft
Unemployed
\end{minipage} & \begin{minipage}[b]{\linewidth}\raggedleft
Prof.~Occup.
\end{minipage} & \begin{minipage}[b]{\linewidth}\raggedleft
Gentrification
\end{minipage} & \begin{minipage}[b]{\linewidth}\raggedleft
Decline
\end{minipage} & \begin{minipage}[b]{\linewidth}\raggedleft
Differential
\end{minipage} \\
\midrule\noalign{}
\endhead
\bottomrule\noalign{}
\endlastfoot
House Price & 1.000 & 0.010 & 0.029 & 0.038 & -0.015 & -0.032 & -0.039 &
0.035 \\
Res. Mob. & 0.010 & 1.000 & -0.288 & 0.187 & -0.016 & -0.043 & -0.050 &
-0.004 \\
Non White & 0.029 & -0.288 & 1.000 & -0.037 & 0.216 & -0.068 & -0.042 &
-0.045 \\
Unemployed & 0.038 & 0.187 & -0.037 & 1.000 & -0.177 & -0.018 & -0.056 &
0.070 \\
Prof.~Occup. & -0.015 & -0.016 & 0.216 & -0.177 & 1.000 & -0.269 &
-0.164 & -0.309 \\
Gentrification & -0.032 & -0.043 & -0.068 & -0.018 & -0.269 & 1.000 &
0.896 & 0.430 \\
Decline & -0.039 & -0.050 & -0.042 & -0.056 & -0.164 & 0.896 & 1.000 &
0.108 \\
Differential & 0.035 & -0.004 & -0.045 & 0.070 & -0.309 & 0.430 & 0.108
& 1.000 \\
\end{longtable}

\subsection{Summary}\label{summary}

The 3 SMD analyses - STM with prevalence covariates, LLM based
neighbourhood characterisation and text embeddings - generated LSOA
level measures that could be linked to the LSOA demographic and house
price change attributes. These included 9 STM topics, 5 LLM
neighbourhood characterisation scores and 3 text embedding scores. These
were linked to the 5 changes in socio-economic data over LSOAs and
combined into a single dataset for use in the cluster analyses described
in the next section.

\section{Neighbourhood Analysis}\label{neighbourhood-analysis}

\subsection{Cluster Analysis}\label{cluster-analysis}

A cluster analysis was undertaken using a model-based clustering
approach to identify groups of observations with similar multivariate
characteristics. The clustering variables were rescaled to zero mean and
unit variance prior to analysis. A series of mixture models were fitted
using the \texttt{mclust} R package (Scrucca et al. 2023). The
clustering approach assumed that the observations arise from a finite
mixture of multivariate Gaussian distributions. For an observation
\(\mathbf{x}_i\), the probability density under a model with \(G\)
components can be expressed as

\begin{equation}
f(\mathbf{x}_i) =
\sum_{g=1}^{G}
\pi_g
\phi(\mathbf{x}_i \mid \boldsymbol{\mu}_g,\boldsymbol{\Sigma}_g)
\label{clus1}
\end{equation}

where \(\pi_g\) is the mixing proportion of cluster \(g\), with

\begin{equation}
\sum_{g=1}^{G}\pi_g=1
\label{clus2}
\end{equation}

\(\boldsymbol{\mu}_g\) is the mean vector for cluster \(g\),
\(\boldsymbol{\Sigma}_g\) is its covariance matrix, and \(\phi(\cdot)\)
denotes the multivariate Gaussian density.

The parameters of each mixture model were estimated using the
expectation-maximisation (EM) algorithm. During the E-step, the
posterior probability that observation \(i\) belongs to cluster \(g\) is
calculated as

\begin{equation}
\tau_{ig} =
\frac{
\pi_g \phi(\mathbf{x}_i\mid\boldsymbol{\mu}_g,\boldsymbol{\Sigma}_g)
}{
\sum_{h=1}^{G}
\pi_h \phi(\mathbf{x}_i\mid\boldsymbol{\mu}_h,\boldsymbol{\Sigma}_h)
}
\label{clus3}
\end{equation}

These posterior probabilities provide a probabilistic measure of cluster
membership. Observations were assigned to the cluster for which they had
the largest posterior probability, producing the final classification
used in subsequent analyses.

Models containing between 1 and 15 mixture components
(\(G=1,\ldots,15\)) were considered. In addition to varying the number
of components, different covariance parameterisations were evaluated,
allowing clusters to differ in volume, shape and orientation. Models
were compared using the Bayesian Information Criterion (BIC):

\begin{equation}
\mathrm{BIC}=2\ell(\hat{\boldsymbol{\theta}})-k\log(n),
\label{clus4}
\end{equation}

where \(\ell(\hat{\boldsymbol{\theta}})\) is the maximised
log-likelihood, \(k\) is the number of estimated parameters and \(n\) is
the number of observations. Larger BIC values indicate a preferable
balance between model fit and model complexity.

The best-fitting model was an EVE model with six clusters. The EVE
covariance structure constrains cluster volume and orientation to be
equal across components while allowing the shape of the covariance
ellipsoids to vary between clusters. The selected model had a
log-likelihood of \(-12696\), included 495 estimated parameters, and had
a BIC of \(-28691\). The next two best-fitting models were also EVE
models, with five and four clusters respectively and BIC values of
\(-29641\) and \(-31446\). The six clusters contained 312, 140, 114,
113, 24 and 83 observations, respectively. The mean posterior
probability of cluster membership and the proportion of observations
with posterior probability \textgreater{} 0.90 are shown in Table
\ref{tab:cluster_props}, indicating the relatively high separability of
the clusters.

\begin{figure}

{\centering \includegraphics[width=1\linewidth]{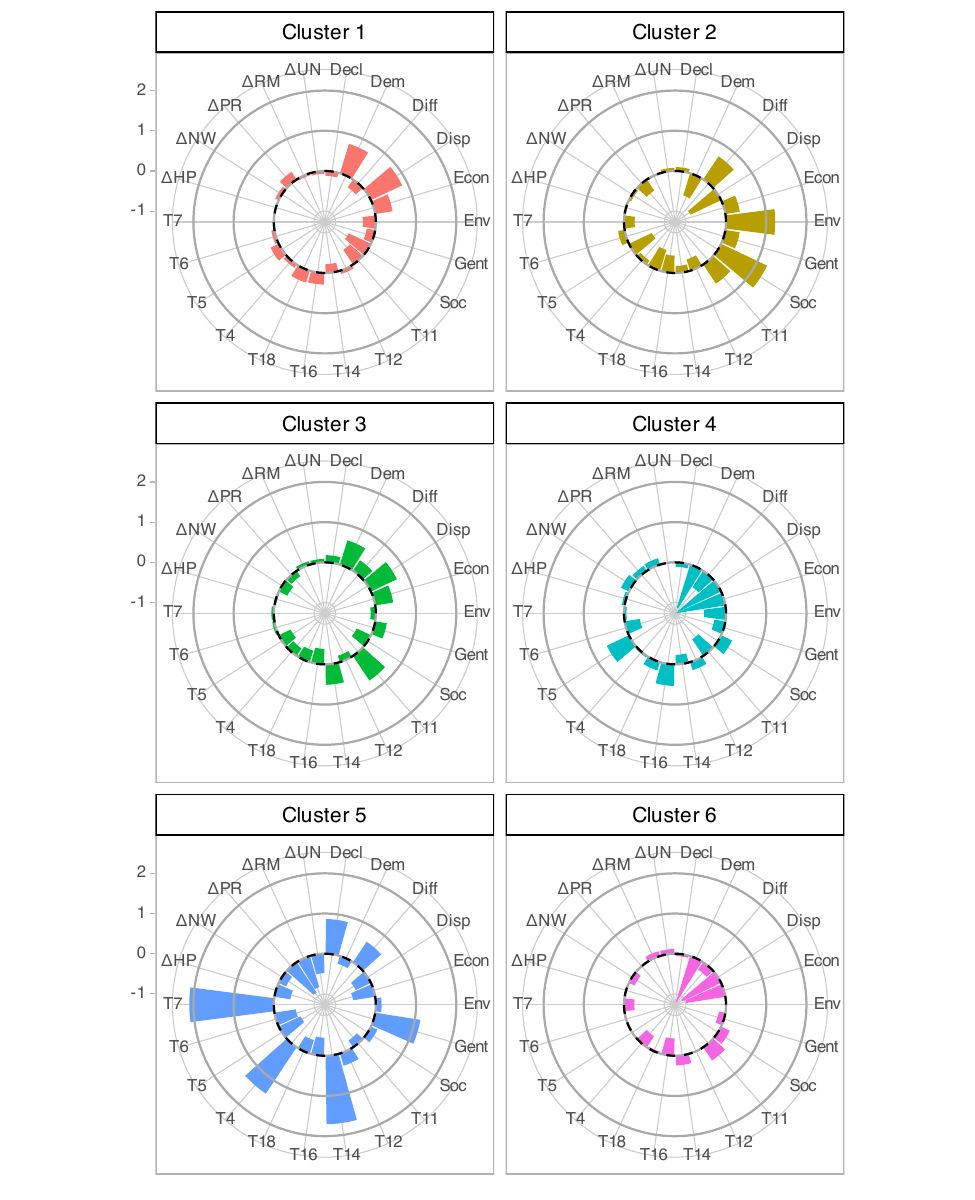} 

}

\caption{The composition (means of the rescaled data) of each cluster.}\label{fig:plot_polar}
\end{figure}

The composition of the clusters can be investigated. Figure
\ref{fig:plot_polar} shows the cluster means of the rescaled data and
indicates that the clusters are capturing distinct combinations of
cultural discourse, community activity, neighbourhood change and
economic/physical characteristics, rather than simply different levels
of house-price change. Table \ref{tab:cluster_summary} suggests some
names that emphasise their dominant signature. These are expanded upon
below.

\textbf{Cluster 1} ``Everyday Urban Flux'' (n = 312). This is the
largest cluster with values close to the average for every
socio-economic slope (ΔHP +0.01, ΔRM --0.05, ΔNW +0.09, ΔUN --0.07, ΔPR
+0.23) indicating a lack of any strong structural trajectory. These are
ordinary, slowly professionalising neighbourhoods. The social media
signature is dominated by generic, youth-leaning chatter (T18 General
Personal \& Family Chat +0.31, T16 Youth \& Student Chat +0.29, T5
Casual Youth Chatter +0.22) lacks rural/heritage discourse (T11 --0.44).
As such, this cluster appears to be ordinary urban, not village or
heritage, territory. The nuance is in the LLM scores: despite flat
demographic changes, perceived Displacement pressure (Disp +0.86) and
Demographic change (Dem +0.76) are elevated, while Social cohesion (Soc
--0.61) and Physical environment (Env --0.32) are below average. These
areas lack measurable change but the social-media discourse around them
already carries an undertone of strain and churn.

\textbf{Cluster 2} ``Settled Affluent Rural/Heritage Enclaves'' (n =
140). This is the most ``stable and comfortable'' cluster. Perceived
Social cohesion (Soc +1.26) and Physical environment (Env +1.22) are the
highest of any cluster, Displacement pressure (Disp --0.84) and
Demographic change (Dem --0.58) are the lowest, and the embedding
Differential score (Diff +0.69) is the highest in the whole typology.
This indicates that the semantic content of the tweets is the strongest
``gentrification/affluent'' indicator out of all six clusters. The
discourse is dominated by T11 Rural Yorkshire Leisure \& Heritage
(+0.57), with generic chatter suppressed (T5 --0.63, T18 --0.54, T16
--0.43). The cluster has a counter-intuitive negative ΔPR (--0.33), the
most negative professionalisation slope outside of Cluster 5. As these
areas have the highest social cohesion and built-environment scores,
this most likely reflects a ceiling effect (already highly
professionalised, little room to rise further) rather than decline.

\textbf{Cluster 3} ``Heritage Villages in Quiet Transition'' (n = 114).
This is a variant on the heritage theme (Cluster 2) but with some
structural and discursive distinctiveness. T14 Pennine Village Heritage
\& Food (+0.51) and T11 Rural Leisure (+0.74) are both high, yet ΔPR is
falling (--0.16) and ΔNW is the most negative of any non-extreme cluster
(--0.24), indicating slower ethnic diversification than elsewhere. The
nuance is that, unlike Cluster 2, the LLM scores here indicate rising
Displacement pressure (Disp +0.71) and Demographic change (Dem +0.63)
with positive Economic activity (Econ +0.45) and embedding Differential
(Diff +0.32), while Social cohesion is below average (Soc --0.39). This
potentially reflects heritage villages experiencing lifestyle- or
amenity-driven pressure (incomers, tourism, food culture), without
ethnic-composition change, that is being picked up in the social media
discourse as disruptive to the cluster's settled character.

\textbf{Cluster 4} ``Diversifying Quietly'' (n = 113). Structurally,
this is a classic fast-turnover neighbourhood profile: every
socio-economic slope is positive and ΔRM (+0.17) and ΔNW (+0.22) are the
highest of any cluster (with Cluster 5 at the opposite extreme) with
increasing mobility, rising ethnic diversity, rising house prices and
professionalisation. The interesting nuance is that the SMD-derived
measures tell almost the opposite story: Economic activity (Econ
--1.27), Displacement pressure (Disp --1.18) and Demographic change (Dem
--1.12) are the most or near-most negative of any cluster. The embedding
Differential (Diff --0.48) is the most decline-leaning in the typology.
In terms of the topics, youth/student chatter dominates (T5 +0.61, T16
+0.54) with rural/heritage and gratitude topics suppressed. This
captures an interesting tension with fast demographic/economic turnover
that is invisible, or even reads as ``decline'', in the social media
discourse, potentially indicating change happening beneath the surface
of local on-line chatter.

\textbf{Cluster 5} ``Settled British-Pakistani Community Core'' (n =
24). This is the best defined and separated cluster with the highest
posterior probabilities in the model summary. Every socio-economic slope
is strongly negative --- ΔHP --0.40, ΔRM --0.81 (the largest magnitude
in the table), ΔUN --0.48, ΔPR --0.90 --- indicating a settled,
low-turnover population with little relative house-price movement and
little professionalisation or mobility change. The topic signature is
unambiguous: T7 Bradford British-Pakistani Community Life is at +2.10,
by far the single largest value anywhere, alongside T14 Pennine
Heritage/Food (+1.70) and T4 UK Political \& Religious Debate (+1.38),
while all the generic chatter topics are suppressed. Embedding scores
are high on both archetypes at once (Gent +1.14, Decl +0.86), which adds
semantic richness to the narrative for this cluster given the strong
negative socio-economic changes. This is an anchored, place-specific,
low-turnover community rather than one undergoing transformation in
either direction.

\textbf{Cluster 6} ``Quiet Margins'' (n = 83). This has a modest churn
profile with ΔRM (+0.13) and ΔUN (+0.12) both mildly positive, ΔNW
negative (--0.15) and ΔHP and ΔPR both flat. What is noteworthy is the
SMD discourse: Demographic change is the most negative of any cluster
(Dem --1.19) with Economic vitality (Econ --0.98) and Displacement
pressure (Disp --1.07) also suppressed, against mildly positive Social
cohesion (Soc +0.24). There is a modest rural/heritage presence (T11
+0.40, T14 +0.24) but much smaller than Clusters 2, 3 and 5. The
combination of rising unemployment/mobility with the weakest
LLM-detected change signal (lowest Econ, Disp and Dem) suggests
low-profile neighbourhoods experiencing a mild economic drift that is
not being discussed or noticed on-line and indicative of areas at risk
of under-the-radar strain rather than visible decline.

\begin{longtable}[]{@{}lrr@{}}
\caption{\label{tab:cluster_props}The mean posterior probability and
proportion of each cluster with a posterior probability
\textgreater0.90.}\tabularnewline
\toprule\noalign{}
Cluster & Proportion \textgreater0.9 & Mean posterior \\
\midrule\noalign{}
\endfirsthead
\toprule\noalign{}
Cluster & Proportion \textgreater0.9 & Mean posterior \\
\midrule\noalign{}
\endhead
\bottomrule\noalign{}
\endlastfoot
1 & 0.9776 & 0.9930 \\
2 & 0.9286 & 0.9740 \\
3 & 0.9211 & 0.9753 \\
4 & 0.9735 & 0.9888 \\
5 & 1.0000 & 0.9993 \\
6 & 0.9157 & 0.9703 \\
\end{longtable}

\begin{longtable}[t]{>{\raggedleft\arraybackslash}p{0.7cm}>{\raggedright\arraybackslash}p{3cm}>{\raggedright\arraybackslash}p{12cm}}
\caption{\label{tab:unnamed-chunk-7}\label{tab:cluster_summary}The names allocated to each cluster and their characteristics based on their dominant components.}\\
\toprule
Cluster & Name & Main Characteristics\\
\midrule
1 & Everyday Urban Flux & Average socio-economic trajectories; elevated youth/family chatter; high perceived displacement pressure and demographic change despite low social cohesion and physical environment\\
2 & Settled Affluent Rural/Heritage Enclaves & High community cohesion and physical environment; strong rural/heritage discourse; lowest displacement pressure and demographic change; declining professional occupation from an already high base\\
3 & Heritage Villages in Quiet Transition & Strong rural/heritage discourse with declining professionalisation and slower ethnic-composition change; elevated perceived displacement pressure and demographic change despite below-average community cohesion\\
4 & Diversifying Quietly & Fast house price rises, mobility, ethnic diversity and professionalisation; dominated by youth/student chatter; lowest perceived economic vitality, displacement pressure and demographic change, with a decline-leaning semantic signal\\
5 & Settled British-Pakistani Community Core & Dominated by Bradford British-Pakistani community and religious/political discourse; largest declines in mobility, professionalisation and unemployment; low-turnover, anchored community signature\\
\addlinespace
6 & Quiet Margins & Mild rises in unemployment and mobility with declining ethnic-composition change; lowest perceived economic vitality and demographic change despite mildly positive community cohesion\\
\bottomrule
\end{longtable}

\begin{figure}

{\centering \includegraphics[width=1\linewidth]{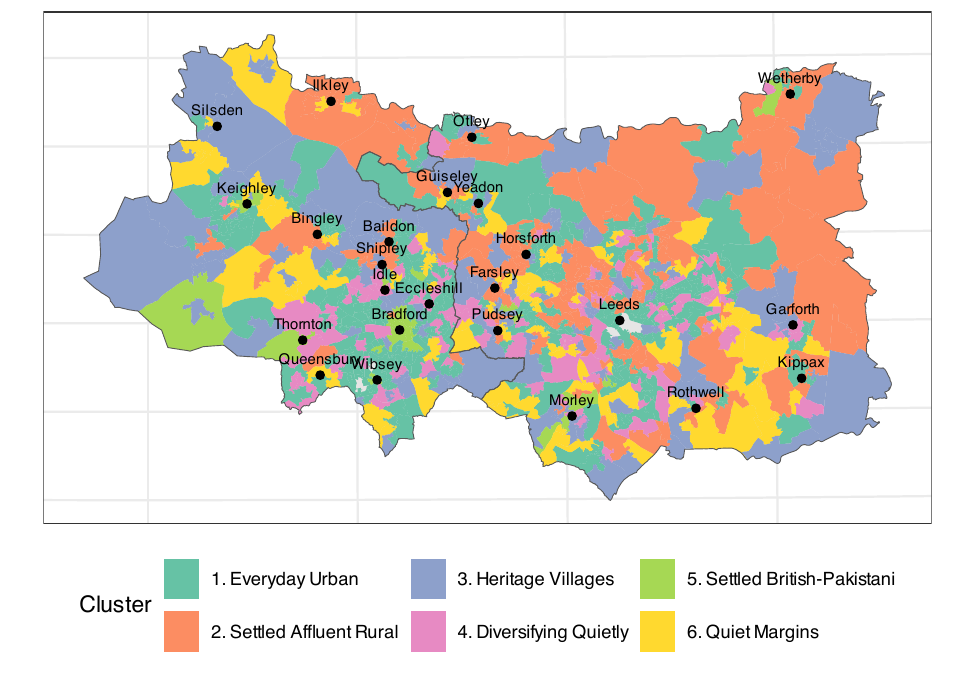} 

}

\caption{The spatial distribution of the neighbourhood clusters, with local context.}\label{fig:plot_clus}
\end{figure}

The size of the effect of each variable used in the clustering is
included in Supplementary materials (Table \ref{tab:eta_tab}). Finally,
the spatial distribution of the clusters can be mapped as in Figure
\ref{fig:plot_clus}.

\subsection{The effects of input variable changes on cluster
membership}\label{the-effects-of-input-variable-changes-on-cluster-membership}

A final analysis is to examine the effects on the cluster membership of
changes in the variables used to construct the clusters. Here the
approach taken is to construct a Generalized Additive Model (GAM). GAMs
(Hastie and Tibshirani 1987; Hastie and Tibshirani 1990) use smooth
functions to capture non-linear relationships. These fit curves through
the data under the assumption that, overall, the fitted curve describes
some related process. Smooths are constructed from a series of basis
functions which allow GAMs to capture complex (non-linear)
relationships. If GAM smooths include observation location as well as
predictor variables, then they can be used capture spatially varying
relationships (Fan and Huang 2022; Comber, Harris, and Brunsdon 2024;
Comber, Harris, Murakami, et al. 2024). In this case, an initial
multinomial regression model was run to determine which of the predictor
variables were dominant and then these were incorporated into GAM
smooths. The dominant predictor variables were the LLM derived Economic
vitality and Social cohesion, and the STM topics Topic 14, Topic 11 and
Topic 7. The GAM was thus specified as follows:

\begin{equation}
\log\frac{P(Y_i = k)}{P(Y_i = 1)} = \eta_{ik} \quad \text{for } k = 2, 3, 4, 5, 6
\label{gam1}
\end{equation}

\begin{equation}
\eta_{ik} = \beta_{0k} + \sum_{j \in \mathcal{V}} f_{jk}(X_i, Y_i) \cdot v_{ij}
\label{gam2}
\end{equation}

\begin{equation}
\eta_{ik} = \beta_{0k} 
  + f_{\text{ev},k}(X_i, Y_i) \cdot \text{ev}_i 
  + f_{\text{t14},k}(X_i, Y_i) \cdot \text{t14}_i \\
  + f_{\text{t11},k}(X_i, Y_i) \cdot \text{t11}_i 
  + f_{\text{t7},k}(X_i, Y_i) \cdot \text{t7}_i 
  + f_{\text{sc},k}(X_i, Y_i) \cdot \text{sc}_i
\label{gam3}
\end{equation}

where \(Y_i\) is the cluster membership (1, 2, 3, 4, 5, or 6),
\(\beta_{0k}\) is the intercept for cluster \(k\) (reference is cluster
1, so no \(\eta_{i1}\)), \(f_{jk}(X, Y)\) is a thin-plate spline basis
function that varies by both space \((X, Y)\) and cluster \(k\),
\(v_{ij}\) are the predictor variables: \(ev\) = Economic vitality,
\(t14\) = Topic 14, \(t11\) = Topic 11, \(t7\) = Topic 7, \(sc\) =
Social cohesion. Note that this estimates probabilities relative to a
reference class, which in this case is cluster 1.

It is possible to examine (and map) observation-level changes in
multinomial membership probabilities arising from input variable changes
from the spatially varying coefficient GAM (Comber, Harris, and Brunsdon
2024) as an extended application of the observed-value discrete-change
approach (Scott Long 1997; Hanmer and Ozan Kalkan 2013), with intervals
following King et al. (2000). The basic approach is to compute the
discrete change in predicted cluster probabilities when a factor used in
the cluster construction (a predictor variable in the GAM) increases by
one standard deviation (+1 SD) with the others held fixed. The effects
for each observation (LSOA) can be calculated to generated unit level
contrasts and then aggregated to generated average contrasts or average
discrete changes (Arel-Bundock et al. 2024).

\begin{figure}

{\centering \includegraphics[width=1\linewidth]{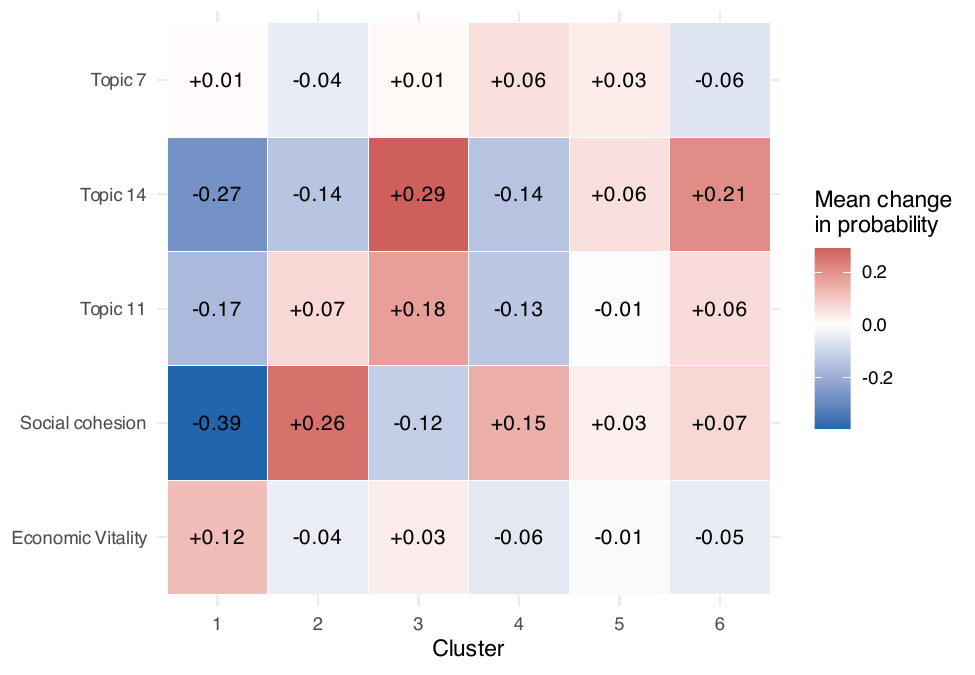} 

}

\caption{The average effect of +1 SD increase in each factor on cluster membership}\label{fig:fig_sd1}
\end{figure}

The results in Figure \ref{fig:fig_sd1} indicate which factor dominates
each cluster, and how strongly the membership of that cluster responds
to the changes in that factor. The figure indicates that, on average, a
+1 SD increase in Social cohesion reduces the probability of membership
of cluster 1 by 0.39 and increases the probability of cluster 2 by 0.26.
Topic 14 had the next largest effect, in this case on clusters 3 and 6.
Changes in Economic vitality have small average effects and changes in
Topic 7 have weak effects across all clusters. In essence, these reflect
the GAM and the degree to which the smooths shrink out the effects of
the predictor variables on the multinomial response.

\begin{figure}

{\centering \includegraphics[width=1\linewidth]{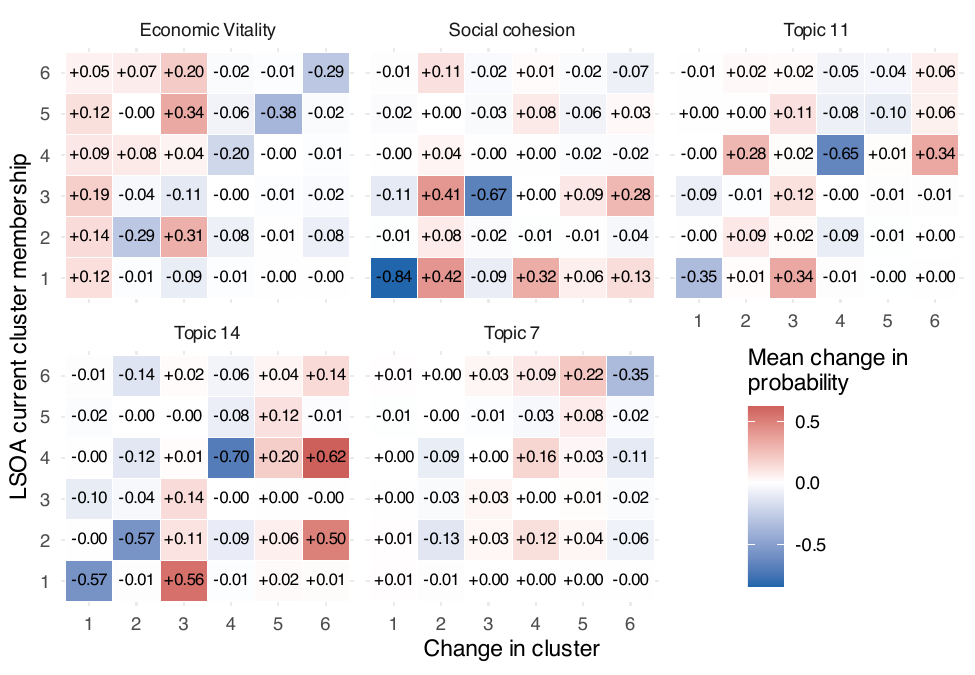} 

}

\caption{The overall effect of each factor on each cluster.}\label{fig:fig_sd1_2}
\end{figure}

However, the averages in Figure \ref{fig:fig_sd1} can mask important
local effects and differences. As a hypothetical example, a change in a
factor might pull some LSOAs out of cluster 1 and others out of cluster
2. It is possible to look at the changes in cluster membership
probability as a result of a +1 SD change in each factor as in Figure
\ref{fig:fig_sd1_2}. Here each cell is the average change in
probability, across areas in that cluster (row), when the factor goes up
by 1 SD and indicates the probability of change to another cluster
(column).

If the diagonal values are considered then negative values indicates a
weakening of current cluster membership and any positive values in that
row indicate the direction of change. Any large negative value indicates
that this factor is a defining difference between the current cluster
and destination cluster. Similarly, positive diagonal values indicate
that the change in the factor reinforces cluster membership
(e.g.~cluster 1 for Economic vitality and cluster 6 for Topic 14). Rows
composed of values near to 0 indicate that the factor has little impact
on cluster membership.

Considering each factor, the following changes are suggested following a
+1 SD change in the factor, noting that these are not causal effects,
they simply describe how the clusters are separated:

\begin{itemize}
\tightlist
\item
  Social cohesion: Cluster 1 areas change to clusters 2 (+0.42) and 3
  (+0.32). Cluster 3 areas lose 0.67, going to clusters 2 (+0.41) and 6
  (+0.28). Other rows barely change. Social cohesion is mainly what
  separates clusters 1 and 3 from clusters 2, 4 and 6.
\item
  Economic vitality: Clusters 2, 5 and 6 shift mostly to cluster 3 and
  cluster 4 shifts to clusters 1 and 2. Cluster 3 loses a little to
  cluster 1 and cluster 1 is reinforced. Cluster 1 is the most
  economically vital, followed by cluster 2.
\item
  Topic 14: Clusters 1, 2 and 4 each lose. Cluster 1 goes to cluster 3
  and clusters 2 and 4 go to cluster 6. Clusters 3, 5 and 6 are
  reinforced. Topic 14 separates cluster 1 from cluster 3, and clusters
  2 and 4 from cluster 6.
\item
  Topic 11: Cluster 4 goes to clusters 2 and 6 and cluster 1 goes to
  cluster 3. Topic 11 helps to separate cluster 1 from cluster 3.
\item
  Topic 7: Cluster 6 goes mainly to cluster 5 (+0.22) and cluster 2
  loses a little to cluster 4. Cluster 4 is reinforced. This is a weaker
  factor that mainly separates cluster 6 from cluster 5, with cluster 4
  relatively high.
\end{itemize}

\begin{figure}

{\centering \includegraphics[width=1\linewidth]{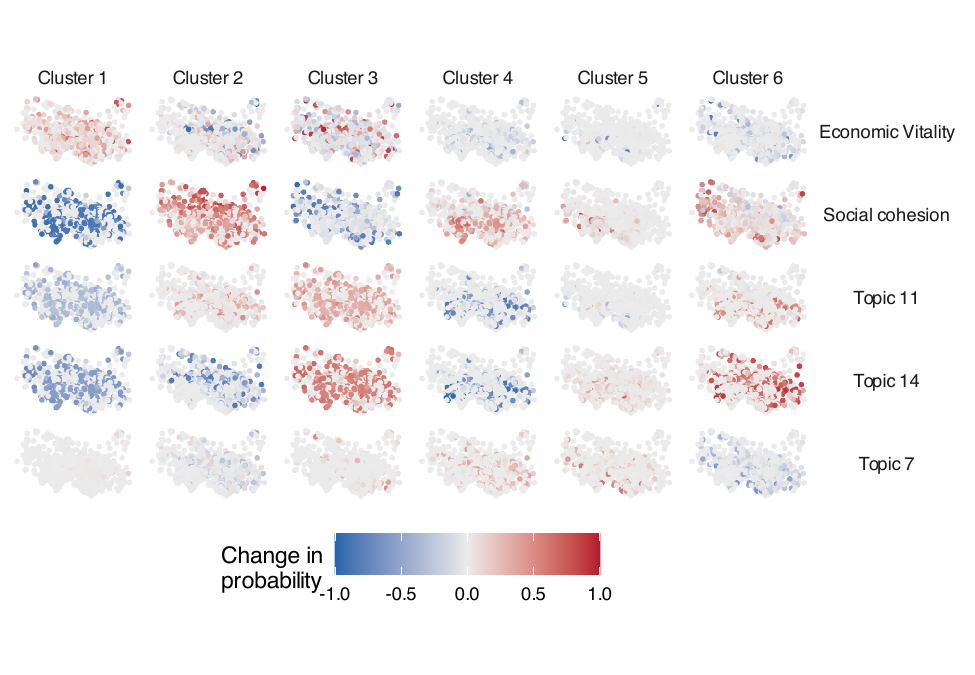} 

}

\caption{The effects on cluster membership as a result of a +1 standard deviation change in each factor, for each cluster, for each LSOA.}\label{fig:fig_sd1_3}
\end{figure}

Finally, the spatial distribution of the effects of changes in factors
on cluster membership for individual LSOAs can be mapped. This is done
in Figure \ref{fig:fig_sd1_3} using points (LSOA geometric centroids).

\section{Discussion}\label{discussion}

This paper sought to unpick the different facets associated with
neighbourhoods and neighbourhood change. It adopted a multidimensional
approach that combined house price and demographic data with the results
of different SMD analyses:

\begin{enumerate}
\def\labelenumi{\arabic{enumi}.}
\tightlist
\item
  Structural Topic Models (STM) with prevalence covariates were used to
  identify topics directly from the SMD corpus. The STM framework allows
  topic proportions to vary with document-level metadata, testing
  whether topics were differentially prevalent in, for example, emergent
  versus declining neighbourhoods.
\item
  LLM based neighbourhood characterisation was used to score tweets
  aggregated to the LSOA by year level across five dimensions related to
  gentrification theory.
\item
  Text embeddings were used to estimate semantic distances to archetypes
  of gentrification and decline as a complement to the topic models and
  LLM characterisation.
\end{enumerate}

Having generated the above SMD derived metrics for each LSOA in the
Leeds and Bradford case study area, two complementary analyses were
undertaken. First, a model-based clustering approach was undertaken to
generate a typology that captured the different types and manifestations
of neighbourhood dynamics. Second, a spatially varying multinomial GAM
was created to explore the spatial variation in the relationship between
cluster membership and different factors.

The individual clusters exhibited very different socio-economic and SMD
derived trajectories suggesting that socio-economic changes and changes
in the social/cultural meanings and activities associated with places do
not necessarily move together. The clusters capture places where local
economic profiles are changing differently from surrounding
neighbourhood, the demographic data capture shifts in neighbourhood
composition and the SMD analyses extracted traces of what people talk
about: their cultural interests, events and consumption, political and
religious discourse and local social and environmental perceptions as
well as semantic associations with gentrification and decline. The
desire to characterise neighbourhoods through socio-economic
transformations and changes in the social, cultural and symbolic
character of those places connects to research that has argued that
neighbourhood change is simultaneously economic, social, cultural and
spatial rather than reducible to property values (Cheshire et al. 2019),
and to cultural capital in which the cultural practices and identities
associated with places are implicated in processes of neighbourhood
(upwards) transformation (Bridge 2006). The analysis demonstrated here
extends further machine-learning and predictive approaches that fuse
multiple heterogeneous data streams (Reades et al. 2019; Gray et al.
2023b) beyond traditional data sources (Chapple et al. 2022).

Potentially the most policy-relevant aspect of the cluster typology is
the repeated divergence between the hard demographic, housing change
trajectories and the SMD discourse. The strongest example of this is
Cluster 4 (``Diversifying Quietly'') in which each structural measures
indicates classic gentrification and turnover: increasing relative house
prices, mobility, ethnic diversity and professionalisation. However the
LLM derived measures relating to economic vitality, displacement
pressure and demographic change are the lowest of any cluster, and the
embedding differential score is the most decline leaning. Cluster 1
(``Everyday Urban Flux'') is almost a mirror of the pattern but at a
lower intensity. It exhibits flat structural changes but elevated LLM
generated perceived displacement pressure and demographic change. This
exemplifies Status--Quality Trade-Off theory (Huu Phe and Wakely 2000)
which describes the interaction of conventional housing / demographic
indicators and the intangible, discursively constructed sense of
neighbourhood character. It is potentially indicative of the early stage
(stealth) phase of gentrification in the rent-gap tradition (Smith 1979;
Hackworth and Smith 2001) in which capital reinvestment and demographic
turnover precede visible changes in the symbolic economy of a place
(Zukin 2009) or any shift in how residents or commentators talk about
it. It is likely that classic administrative and structural indicators
(e.g.~house price and deprivation indices), and softer discursive
indicators alone will not provide sufficient information to support
proactive interventions (e.g.~inclusionary zoning, pre-emptive
right-to-return schemes, etc.). and structural indicators may lag the
underlying process.

Clusters 2 (``Settled Affluent Rural'') and 3 (``Heritage Villages'')
both show classic signatures of rural/village gentrification, associated
with incomer-driven lifestyle and amenity consumption overlaid onto an
already-stable population (Phillips 1993, 2004) but with an important
difference. Cluster 2 appears to be a stable endpoint, with high social
cohesion and physical environment and low displacement pressure, while
Cluster 3 has elevated LLM displacement pressure and demographic change
scores alongside negative social cohesion scores. This suggests that the
heritage villages in this cluster are mid-process, experiencing the
social friction of incomer/tourism pressure without yet showing up as
compositional (ethnic) change, consistent with the concept of ``elective
belonging'' (Savage et al. 2005) in which middle class residents
construct a sense of ownership over a place's heritage and character,
that can be in tension with newer arrivals, and with middle-class
neighbourhood ``colonisation'' operating through social and cultural
capital rather than simple price displacement ({Butler et al.} 2003).
Cluster 3 is a candidate for heritage or tourism related interventions
such as second-home or short-let restrictions and conservation area
policies, since the pressure appears to be amenity- and visitor-driven
rather than price-driven.

Cluster 5 (``Settled British-Pakistani'') is potentially the most policy
sensitive group. All of the structural slopes are strongly negative,
indicating a lack of change, and topics are dominated by
British-Pakistani community discourse. It would be naive to interpret
this through a `no change' lens as indicative of decline or stagnation,
as ethnic clustering is commonly associated with low residential
turnover with strong internal ties that reflect strong, cohesive
communities rather than enforced segregation (Peach 2009; Simpson 2007;
Phillips 2006). These neighbourhoods are settled and connected rather
than segregated and failing (Simpson et al. 2009), suggesting the need
for neighbourhood typologies that recognise stability as a legitimate
form of neighbourhood development. In multi-ethnic urban areas,
established communities may remain socially and culturally stable
without experiencing significant gentrification or decline. Policies
designed primarily to detect and respond to change may therefore
overlook these neighbourhoods or mistakenly interpret strong cultural,
religious or political activity as a sign of social tension instead of a
well established and cohesive community.

Cluster 6 (``Quiet Margins'') includes neighbourhoods that are
accumulating some structural strain such as mild rises in unemployment
and mobility while generating almost no on-line discourse about it (low
perceived economic vitality, displacement pressure, demographic change,
etc.). This raises a methodological point with direct policy
consequences: a low SMD signal is ambiguous. It may be because nothing
is happening or because users are not posting about it. Additionally,
geo-tagged social media users are not a representative sample (Hecht and
Stephens 2014; Johnson et al. 2017; Shelton et al. 2015; Longley et al.
2015). The quiet discourse of Clusters 1 and 6 may reflect who does not
post rather than what is not happening, in which case these clusters are
where digital traces from SMD will fail and potentially the
neighbourhoods that most need attention, reflecting wider critique of
the use of big data in such studies generally (Shelton et al. 2015). As
such, SMD-derived indicators should be treated with caution as they
privilege digitally visible populations and neighbourhoods with
combinations of structural changes and low SMD discourse (e.g.~Cluster
6) may reflect under-detection rather than stability.

When considered together, the six clusters support the central argument
that neighbourhood change is multidimensional and that its economic,
demographic and socio-cultural facets do not move together (Cheshire et
al. 2019). However, it is possible to generate a specific, policy
friendly typology that identifies genuine rapid structural changes that
may be discursively invisible (Cluster 4) or muted (Cluster 1, Cluster
6), that distinguish between stable affluence (Cluster 2) and heritage
areas under amenity pressure (Cluster 3) and adds nuance to a settled
minority-ethnic community that a generic structural change based
framework would mis-characterise (Cluster 5). In this context SMD
derived measures can be used to independently inform on neighbourhood
character rather merely serve as a check on structural indicators, and
could be used to provide bias-correction to commonly used composite
gentrification or displacement indices (e.g. Flinders and Adam (2025);
Almeida (2021)). The results also indicate the potential advantages of
monitoring rapid but potentially hidden structural turnover, amenity and
heritage related pressure (through tourism or conservation tools),
settled minority-ethnic stability (to avoid being pathologised as
declining or segregated), and low level strain (via a presumption of
under-detection rather than an absence of need).

\section{Conclusion}\label{conclusion}

This paper demonstrates a data driven approach for characterising
neighbourhood change as a multidimensional process rather than as a
single continuum of gentrification, improvement or decline. It
integrated relative house price change, time series demographic data and
SMD traces to construct a typology of neighbourhood change in Leeds and
Bradford. The SMD were analysed through STM, LLM based neighbourhood
characterisation and text embeddings, and a spatially varying
multinomial GAM was used to examine how changes in the inputs affect
cluster membership. The analysis did not build indices or regression
models of change, since SMD are sparse, noisy and largely uncorrelated
with socio-economic data. Instead, it treated the SMD derived measures
as multiple, partial indicators of neighbourhood character. The
resulting clusters showed that the economic, demographic and
socio-cultural facets of neighbourhood change do not necessarily move
together. In some neighbourhoods, rapid structural turnover was
discursively invisible (Cluster 4) or muted (Clusters 1 and 6). Heritage
villages were experiencing amenity and incomer related pressure that was
not evident in their ethnic composition (Cluster 3). A settled
British-Pakistani community showed a stability that a change based
framework would potentially mis-characterise as stagnation or decline
(Cluster 5). In this way, the approach provides a means of quantifying
some of the less tangible dimensions of neighbourhood character, and of
disentangling the different manifestations of neighbourhood change, at
scale.

There are a number of limitations and areas for further work. The SMD
were highly temporally imbalanced, with the majority of geo-tagged
tweets from 2014 and 2015, and geo-tagged social media users are not a
representative sample of neighbourhood residents. As such, low SMD
signals are ambiguous and may reflect who does not post rather than what
is not happening. The associations identified here are also descriptive
rather than causal. Further work could extend the analysis to other
cities and time periods. Because the Twitter / X API has closed, it
could also draw on other sources of digital traces, such as online
reviews or property listings. It could also examine the use of SMD
derived measures to inform or bias-correct composite gentrification and
displacement indices. Notwithstanding these limitations, the results
suggest that combining structural and discursive indicators can support
a more nuanced and policy relevant understanding of neighbourhood
change. Such an understanding would identify hidden structural turnover
and distinguish stable affluence from heritage areas under amenity
pressure. It would also recognise settled minority-ethnic stability as a
legitimate form of neighbourhood development, and treat low level strain
in areas with little digital visibility as potential under-detection
rather than an absence of need.

\newpage

\section*{Supplementary Material}\label{supplementary-material}
\addcontentsline{toc}{section}{Supplementary Material}

\renewcommand{\thefigure}{S\arabic{figure}}
\setcounter{figure}{0}
\renewcommand{\thetable}{S\arabic{table}}
\setcounter{table}{0}

\subsection*{LLM Interpretation of STM
topics}\label{llm-interpretation-of-stm-topics}
\addcontentsline{toc}{subsection}{LLM Interpretation of STM topics}

The following query was passed to Claude (Anthropic 2026) for an
interpretation of each STM topic and the results are shown in Table
\ref{tab:topic_themes} below.

\begin{Verbatim}[
  breaklines=true,
  breakanywhere=true,
  breaksymbolleft={},
  breaksymbolright={}
]

Query: 
I would like you to generate some Interpretations for topics generated using the `stm` R package. Below I have extracted the top 100 words for the topics that are significantly associated with changes in various socio-economic variables.
The topics of interest are: 
  predictor topics          
  <chr>     <chr>           
1 hp_beta   4               
2 nw_beta   5, 12, 14, 18   
3 pr_beta   6, 7, 11, 16, 18
4 rm_beta   5, 6, 7, 12, 16, 17       
5 un_beta   7   
> 
> Their meaning is as follows
# - hp_beta is relative house price change
# - rm_beta is change in residential mobility
# - nw_beta is change in non-white residents
# - un_beta is change in  unemployment 
# - pr_beta is change in professional occupation

Please could you generate Interpretations for the topics? Please format the result to be a `tribble`, with columns describing the Topic number (`Topic`), a short 3-5 word label (`Label`), an 70-90 word interpretation (`Interpretation`), a 15-20 word theme (`Theme`), a 3-5 word neighbourhood theme (`Neighbourhood Theme`) and a 3-40 word set of caveats (`Caveats`). The 100 words associated with each of the 20 topics are printed below

# STM topics 
tops <- labelTopics(stm_fit, n = 100)
top_words <- tops$prob|> apply(1, function(x) paste(x, collapse = ", "))
top_words
[1,] "drinking, photo, pudsey, ale, beer, ipa, pale, horsforth, amp, good, flavour, brewery, rothwell, light, hop, nice, stout, coffee, woodlesford, roverrambles, like, bar, fruity, fleece, aroma, porter, new, black, bitter, great, really, quite, fruit, hoppy, chocolate, shack, sour, blonde, lovely, american, dark, one, session, rspb, cap, bit, dogwalker, taste, decent, last, citrus, brewdog, pint, collar, cider, getting, saltairebrewery, hops, smooth, dry, beers, night, inn, refreshing, lager, love, red, old, well, flavours, tropical, tattoo, first, sweet, full, bridge, orange, pub, lots, southleeds, another, wine, best, english, citra, golden, bad, malty, subtle, brew, tasty, hall, nmbco, perry, station, yorkshire, easy, brewed, leeds, kirkstallbrew"                                      
[2,] "leeds, park, festival, roundhay, hyde, kirkstall, airport, just, day, temple, leedsfestival, abbey, one, yorkshire, bramham, newsam, time, leedsfest, west, photo, back, last, weekend, posted, amazing, best, leeds…, night, fest, today, love, good, armley, yeadon, officialrandl, international, old, photography, randl, university, another, see, amp, mint, morning, roundhaypark, yesterday, adel, photographer, great, music, fun, photoshoot, like, lbiairport, first, stunning, lake, year, life, get, chapeltown, happy, home, got, campus, stage, beautiful, lovely, way, world, beaver, beckett, works, estate, next, gorgeous, can, lba, people, ever, sun, now, summer, bus, week, years, new, house, templenewsam, sunday, going, luvyorkshire, little, walk, luvthenorth, carnival, take, holiday, well"
[3,] "just, like, one, think, now, good, dont, thats, ive, really, get, can, well, youre, got, know, yes, bit, love, though, hes, time, see, still, ill, never, right, back, much, thing, will, yeah, say, first, want, cant, theres, better, thought, new, look, didnt, theyre, way, great, sure, make, looks, people, even, going, old, probably, stuff, haha, something, said, man, yep, maybe, lot, live, always, isnt, last, rustyb, doesnt, might, nice, take, enough, years, little, quite, seen, david, watching, best, also, pretty, morning, big, music, need, amp, ever, must, sorry, sounds, actually, give, youve, around, yet, album, play, ewanmitchell, may, song, things"
[4,] "amp, will, get, just, can, like, dont, people, good, one, well, think, see, labour, now, know, bbcleeds, need, time, lol, hope, best, cant, vote, next, want, right, make, many, never, back, thats, say, still, last, ukip, going, stop, better, take, bbcqt, much, party, got, great, years, please, agree, another, way, even, said, martinkelner, nhs, look, israel, world, work, day, really, come, money, new, yes, luck, sure, always, tax, lizgreenlive, didnt, leave, wrong, also, must, left, change, first, bbc, cameron, election, believe, may, support, political, made, give, done, read, muslims, country, long, muslim, wont, nothing, gaza, happy, leeds, use, man, keep"
[5,] "just, like, get, dont, cant, haha, good, love, one, now, got, fuck, fucking, can, know, need, think, shit, time, night, going, will, want, day, ive, people, back, well, even, really, today, see, work, ill, tomorrow, yeah, wait, last, feel, happy, much, tonight, hahaha, actually, getting, man, best, thats, come, mate, still, right, life, never, new, ever, bed, sleep, hate, gonna, year, week, next, someone, look, bad, make, mum, always, little, done, say, watching, birthday, lad, nice, youre, better, hes, didnt, home, wanna, watch, way, weekend, days, take, thing, though, stop, first, school, sick, looking, big, said, funny, lol, everyone, wish"
[6,] "thanks, good, great, amp, just, will, day, well, thank, today, hope, one, love, get, can, see, time, now, like, happy, back, new, going, think, lovely, done, nice, looking, know, work, got, dont, much, cant, morning, yes, need, really, thats, night, best, forward, tomorrow, week, last, year, next, look, still, ive, amazing, ill, luck, please, following, youre, tonight, people, weekend, sure, birthday, leeds, soon, way, first, always, looks, follow, wait, home, keep, come, bit, mate, getting, want, make, better, right, though, days, guys, big, many, old, fab, another, beautiful, enjoy, watching, evening, may, long, fantastic, brilliant, help, take, haha, twitter, sorry"
[7,] "bradford, city, day, just, yorkshire, today, time, west, one, love, white, centre, food, photo, allah, shopping, rose, night, new, happy, good, media, back, posted, museum, great, see, birthday, national, morning, like, work, last, alhambra, amp, well, little, pakistan, get, chicken, first, word, bradfordbrewery, bar, theatre, now, ramadan, got, another, family, can, old, done, foodporn, best, fun, will, beautiful, presentation, prizes, house, donate, science, home, made, curry, ppl, life, really, looking, lovely, today’s, road, amazing, know, nice, thank, week, guys, foodie, big, year, look, hai, delicious, cinema, tonight, nandos, odsal, much, christmas, getting, university, ready, thanks, always, way, maryamnsharif, next, tasty"                                                     
[8,] "mate, get, good, just, one, got, like, will, dont, now, think, time, well, game, day, going, see, cant, can, amp, leeds, back, today, haha, know, last, ive, night, thats, pal, still, ill, leedsrhinos, great, firman, tonight, need, tomorrow, next, hes, week, never, though, year, rhinos, come, work, love, best, much, right, hope, look, win, seanrowntree, fuck, done, ever, getting, better, big, didnt, yeah, old, rugby, want, even, team, cheers, fucking, lad, play, new, years, make, youre, gonna, mayne, top, away, jonnyhoward, talbotjordan, season, say, league, watching, home, thought, looking, lilleyste, man, might, way, players, really, rugbyam, days, fans, sure, seen"
[9,] "leeds, amp, headingley, day, today, new, yorkshire, love, just, time, morning, night, last, back, one, happy, great, little, stadium, work, lovely, week, first, get, cricket, gym, good, birthday, tonight, home, allerton, chapel, now, session, west, got, wetherby, well, year, like, done, can, beautiful, christmas, training, coffee, best, made, looking, fitness, another, bit, garforth, working, going, ready, weekend, sunday, see, amazing, thank, getting, friday, much, will, run, ’ve, thanks, house, world, big, still, fun, food, start, yesterday, next, days, tropical, saturday, two, make, way, look, class, lunch, years, always, family, place, finally, come, pizza, evening, really, afternoon, weeks, cake, old, england"                                                                      
[10,] "club, otley, social, brudenell, bingley, leeds, guiseley, amp, night, just, hall, nathbrudenell, yorkshire, music, great, last, day, chevin, house, golf, tonight, live, drinking, new, good, photo, love, bhafc, time, one, guiseleyafc, oulton, gig, harewood, centre, today, see, trainergirl, otleyccsocial, park, bar, back, now, beer, weekend, amazing, cookridge, lovely, daviesa, show, get, like, got, east, best, first, north, west, another, otleycycleclub, thanks, home, rugby, menston, well, nice, big, birthday, beautiful, arrived, week, band, emmerdale, going, little, posted, will, really, happy, town, high, ives, forest, old, venue, fun, alwoodley, sunday, harden, wedding, thank, tour, festival, pub, years, christmas, still, favourite, evening, set"                                    
[11,] "yorkshire, day, ilkley, saltaire, united, today, kingdom, morning, park, walk, just, christmas, happy, beautiful, love, run, time, photo, mill, new, little, moor, shipley, salts, lovely, one, sunday, first, west, home, village, great, family, nature, hall, weekend, canal, baildon, sun, evening, afternoon, back, autumn, running, like, garden, get, good, church, dog, year, posted, woods, snow, nice, amp, spring, golden, work, meanwood, last, bike, farm, winter, tree, station, sunshine, fun, bit, night, old, railway, got, see, now, birthday, best, country, view, middleton, train, drawing, look, valley, ride, westyorkshire, reservoir, river, done, house, bridge, sky, looking, amazing, art, wedding, walking, friends, summer, can"                                                            
[12,] "lufc, leeds, will, good, just, get, mot, road, game, elland, like, now, today, well, one, dont, mate, think, can, day, football, see, back, great, time, fans, team, season, new, hes, come, philhayyep, cant, united, club, got, win, players, last, still, know, cellino, thats, hope, need, amp, tonight, play, lol, first, better, next, waccoe, away, league, going, really, apopey, done, right, best, shit, night, say, player, goal, half, home, please, man, seen, another, take, love, playing, much, big, top, didnt, fuck, lufcstats, happy, fucking, never, old, even, ive, sure, way, want, games, yorkshire, bad, long, years, money, manager, look, year, lets"
[13,] "get, just, bcafc, good, game, like, will, now, amp, well, got, can, one, lol, dont, time, back, win, man, come, great, cant, think, today, team, see, season, hes, going, last, still, need, know, tonight, goal, best, shit, fans, play, league, thats, hope, day, football, big, city, love, better, never, fuck, done, cup, next, world, away, new, lfc, england, take, fucking, players, mate, liverpool, player, even, home, first, right, getting, top, much, happy, way, look, people, say, year, work, arsenal, ive, bad, please, week, watching, playing, chelsea, make, didnt, another, always, united, half, club, watch, games, thanks, lad, looking, seen, want"
[14,] "yorkshire, west, lbc, haworth, morley, thornton, just, james, bronte, photo, posted, katyperry, fat, country, keighley, tea, homemade, like, boston, black, chicken, day, white, served, spa, one, tonight’s, another, thought, today, little, silsden, amp, christmas, garlic, trump, oxenhope, homegrown, salad, people, really, oakworth, delicious, pie, made, last, chips, veggie, katy, humour, time, now, every, says, got, happy, food, onion, green, ’re, get, bath, railway, will, land, british, new, ’ve, recipe, can’t, show, red, love, top, know, bread, racist, can, nawaz, sourdough, radio, two, finally, blog, baking, tomato, west…, check, fresh, roast, lad, arrived, first, don’t, topped, right, brexit, tasty, week, thick"                                                                      
[15,] "follow, please, love, sos, see, much, can, barsandmelody, get, lukesos, like, mean, luke, just, youre, dont, day, omg, cant, cute, back, will, one, calumsos, world, hemmings, ive, hey, michaelsos, tweet, make, know, never, birthday, guys, want, meet, ashtonsos, leeds, come, good, happy, got, niallofficial, now, pls, gonna, think, going, really, time, votesos, georgeujworld, overloadmusic, harrystyles, need, amazing, thanks, night, people, best, tomorrow, jaibrooks, realliampayne, ill, harvey, ilysm, lukebrooks, unionjworld, jamesyammouni, camerondallas, everything, wait, xxx, jjujworld, brooksbeau, ever, actually, nice, mate, boys, even, direction, better, thing, proud, always, still, today, janoskians, pick, fav, joshujworld, feel, hope, say, life, justin, bam, video"               
[16,] "just, like, dont, get, love, know, cant, one, people, can, now, need, time, want, really, life, much, ive, even, day, got, think, feel, going, good, lol, youre, still, today, back, actually, shit, will, gonna, happy, see, fuck, man, make, night, never, ever, someone, last, ill, fucking, thats, come, sleep, always, work, new, right, best, hate, wanna, omg, look, tomorrow, year, bed, way, thing, stop, getting, say, home, well, birthday, girl, uni, wait, days, god, miss, made, didnt, first, something, things, tonight, makes, take, leeds, yeah, though, amp, bad, next, every, better, done, house, girls, haha, [laughing emoji][laughing emoji][laughing emoji], wish, excited, literally, thanks"
[17,] "jesus, lord, love, god, holy, amen, thank, que, name, glory, one, yes, will, pas, youre, every, les, awesome, amp, hes, cest, just, hallelujah, keighley, jai, can, true, christ, des, dont, king, heavens, bless, know, never, good, now, blood, mais, pray, day, get, elle, father, ever, nothing, new, une, night, like, life, follow, wow, work, cant, plus, qui, see, blessed, way, best, back, alone, est, prayer, avec, need, pour, grace, victory, time, [laughing emoji][laughing emoji][laughing emoji], harper, dans, moi, hearts, satan, give, faith, suis, hope, sur, beautiful, want, peace, mercy, let, thanks, full, gods, evil, cross, always, spirit, people, help, happy, great, heart, power"        
[18,] "just, love, get, cant, like, good, day, lol, amp, dont, now, time, can, got, xxx, night, today, well, will, need, one, know, work, back, hope, think, see, haha, happy, going, wait, ive, want, really, much, bed, morning, come, thats, people, new, done, tonight, last, still, feel, home, sleep, tomorrow, look, best, thanks, thank, mum, never, ill, amazing, little, birthday, getting, week, always, better, make, right, please, nice, watching, even, gonna, leeds, lovely, way, yes, say, days, next, wish, looking, bad, long, yeah, hate, year, baby, ever, great, bit, omg, xxxx, hes, life, watch, feeling, big, first, beautiful, take, though, keep"
[19,] "fristaider, annex, leoslegacy, lol, good, hope, just, gunnymark, cbb, glennlayfield, thanks, tracybfc, yeah, will, lovely, debbiefishman, mcs, get, mate, morning, love, day, like, craigstewart, ssleedsutd, mirfchris, dont, charlie, nffc, know, little, well, can, thats, now, got, bbuk, mayhemsoutheast, friend, great, today, djjimmyj, bernie, trigger, cant, nigelsquair, one, xxx, pedsboss, night, time, ooh, think, back, buddy, going, take, sleep, ive, need, naughtyden, better, really, yes, following, much, weekend, welcome, work, family, sorry, bit, thank, see, happy, nice, thankyou, bad, raffles, try, hes, scott, still, always, care, tonight, getting, done, want, people, shes, days, look, poor, soon, feel, bloody, keep, right, bet"                                                      
[20,] "just, day, good, get, night, time, now, today, leeds, one, back, last, tonight, going, cant, will, work, well, like, can, got, see, great, think, home, love, ive, week, new, tomorrow, best, still, need, dont, come, next, first, morning, haha, happy, year, watching, getting, hope, much, game, done, weekend, really, wait, know, ever, days, looking, never, right, better, mate, way, big, birthday, ill, amp, win, excited, bit, thanks, though, christmas, bed, world, two, please, thats, yeah, nice, sure, house, people, long, man, start, made, yes, didnt, want, football, another, feel, watch, team, seen, even, look, half, looks, hours, season, forward, life"

\end{Verbatim}

\begingroup\fontsize{8}{10}\selectfont

\begin{longtable}[t]{r>{\raggedright\arraybackslash}p{2cm}>{\raggedright\arraybackslash}p{9cm}>{\raggedright\arraybackslash}p{3cm}}
\caption{\label{tab:unnamed-chunk-9}\label{tab:topic_themes}LLM generated interpretations of the STM topics that were significantly associated with socio-economic changes, with a summary label and key themes.}\\
\toprule
Topic & Label & LLM Interpretation & Key Themes\\
\midrule
4 & UK Political \& Religious Debate & This topic captures explicitly political and topical debate, dominated by UK party politics (Labour, UKIP, Cameron, election), broadcast media commentary (BBC Question Time, BBC Leeds), and contentious international/religious topics (Israel, Gaza, Muslims). The mix of domestic policy terms (NHS, tax) with foreign-affairs and religious vocabulary suggests users engaging in opinionated, often combative commentary on news events rather than personal or lifestyle content. Its strong negative association with house-price change indicates this politically engaged discourse is disproportionately concentrated in LSOAs where relative house prices are stagnating or declining. & UK party politics, broadcast news commentary, and contested international and religious affairs discussed in an opinionated register\\
5 & Casual Youth Social Chatter & This topic is dominated by informal, expletive-laden conversational language (haha, fuck, mate, gonna) covering everyday personal life — sleep, mood, weekends, school, and social plans. There is no distinctive place-based or thematic vocabulary; it reads as generic, high-frequency social-media banter typical of younger users. Its positive associations with rising non-white population share, rising residential mobility, and rising professional occupation suggest this casual register clusters in LSOAs undergoing faster demographic turnover and socio-economic change, consistent with younger, more transient populations concentrating where neighbourhood change is fastest. & Informal, expletive-heavy everyday chatter with no fixed subject, typical of younger social-media users\\
6 & Gratitude \& Well-Wishing Chatter & This topic is characterised by highly polite, appreciative Twitter-style language — thanks, congratulations, well-wishes, birthdays, and phatic social affirmation ('lovely', 'amazing', 'looking forward'). It lacks substantive topical content, functioning as a register of social bonding and reciprocal courtesy typical of everyday platform interaction. Its positive associations with rising residential mobility and rising unemployment, alongside a negative association with rising professional occupation, suggest this courteous, low-content style is more prevalent in LSOAs experiencing higher turnover and labour-market strain, and less prevalent where the population is professionalising. & Polite, low-content social courtesy and reciprocal well-wishing, not tied to a specific subject\\
7 & Bradford British-Pakistani Community Life & This topic centres on Bradford's British-Pakistani community and city life: food culture (curry, Nandos, foodporn), religious observance (Allah, Ramadan), heritage sites (Alhambra, museum), and family-oriented content. It reads as anchored, place-specific community discourse rather than transient commentary. Its significant negative associations across house-price change, residential mobility, unemployment change, and professional occupation change indicate this discourse is concentrated in LSOAs experiencing below-average change on every one of these measures — consistent with a settled, lower-turnover community rather than one undergoing rapid socio-economic transformation in either direction. & Bradford's British-Pakistani community life, food culture, and religious observance rooted in a stable, low-turnover population\\
11 & Rural Yorkshire Leisure \& Heritage & This topic reflects countryside leisure and heritage tourism around affluent West Yorkshire villages and moorland — Ilkley, Saltaire, Baildon, Shipley — featuring walking, running, gardens, and seasonal imagery. The vocabulary evokes an established, outdoor-oriented lifestyle rather than urban or transient content. Its negative associations with rising non-white population and rising professional occupation suggest this discourse is more prevalent where these measures are changing more slowly than average — plausibly because such areas are already predominantly white and already have high professional-occupation shares, leaving little further room to rise. & Countryside leisure, heritage villages, and outdoor lifestyle content in already-affluent, demographically stable locations\\
\addlinespace
12 & Leeds United Football Fandom & This topic captures Leeds United (LUFC) football fandom — match-day commentary, players, managers, league standings, and fan banter, including strong language typical of sports forums. It is a clearly identifiable, single-subject topic with minimal overlap elsewhere. Its positive associations with rising residential mobility and rising non-white population share suggest football-fandom discourse is disproportionately concentrated in LSOAs undergoing faster demographic turnover, potentially reflecting younger, more mobile, or more diverse populations engaging heavily with local football culture rather than any direct causal link between the sport and neighbourhood change. & Leeds United match-day commentary and fan culture, concentrated in neighbourhoods with faster demographic turnover\\
14 & Pennine Village Heritage \& Food & This topic blends Brontë-country heritage tourism (Haworth, Oxenhope, Oakworth, Keighley) with home cooking and recipes (sourdough, homemade, roast), alongside occasional political commentary (Brexit, Trump). It reads as a heritage-and-domestic-life topic rooted in the Worth Valley's historic Pennine villages. Its significant negative association with rising non-white population share indicates this discourse is concentrated in LSOAs where ethnic composition is changing more slowly than average — consistent with these villages' historically low diversity and limited demographic turnover relative to Leeds and Bradford's more urban, rapidly diversifying neighbourhoods. & Brontë-country heritage tourism, home cooking, and local identity in historically low-diversity Pennine villages\\
16 & Youth \& Student Social Chat & Similar in register to other informal chat topics, this one leans toward student and young-adult life, with explicit references to university ('uni'), emotional expression, sleep, and casual profanity. It lacks distinctive place-based content. Its strong negative association with rising residential mobility, combined with a positive association with rising professional occupation, is a notable and somewhat counter-intuitive pairing — suggesting this discourse is concentrated in LSOAs with comparatively low turnover that are nonetheless professionalising, potentially reflecting settled, upwardly-mobile young-professional or student communities rather than transient, high-churn populations. & Informal student and young-adult chatter concentrated in settled, professionalising neighbourhoods with low turnover\\
17 & Christian Worship \& Prayer & This topic is dominated by Christian devotional vocabulary (Jesus, Lord, God, Amen, Hallelujah, Christ, blessed, grace, faith, prayer, spirit, mercy, cross), reflecting worship and praise discourse. A large share of its remaining words are French function words (que, pas, les, c'est, j'ai, des, mais, elle, qui, avec, pour, dans, moi, suis), suggesting STM has pooled francophone devotional content with English-language worship rather than splitting them apart. Its significant negative association with residential mobility change indicates this discourse concentrates in LSOAs where population turnover rises more slowly than average — consistent with a rooted, stable faith community rather than a transient one. & Christian worship, prayer, and praise discourse spanning English and French, concentrated in stable, lower-turnover faith communities\\
18 & General Personal \& Family Chat & Like several other topics here, this one is dominated by generic, affectively warm personal chat — sleep, family ('mum', 'baby'), well-wishes, and everyday routine — with no strong thematic anchor. Its positive associations with rising non-white population share, rising unemployment, and rising professional occupation form a mixed and somewhat contradictory socio-economic signal. This likely reflects the fact that this style of everyday personal discourse is common across many kinds of changing neighbourhoods, making the topic more a marker of general demographic dynamism than of any single, coherent socio-economic trajectory. & Generic everyday personal and family chatter, weakly linked to several, partly conflicting change measures\\
\bottomrule
\end{longtable}
\endgroup{}

\subsection*{Variable effect}\label{variable-effect}
\addcontentsline{toc}{subsection}{Variable effect}

It is also possible to calculate an effect size measure for every
clustering variable. For each variable, the proportion of its variance
associated with cluster membership was calculated based on the sums of
squared difference with and between clusters:

\begin{equation}
\eta^2 = \frac{SS_{\mathrm{between}}} {SS_{\mathrm{total}}}
\label{eta1}
\end{equation}

Here \(\eta^2\) close to zero indicates that the variable has similar
values across clusters, whereas a larger value indicates that the
variable strongly differentiates the clusters. These are shown in Table
\ref{tab:eta_tab} and indicate that Displacement Pressure, Demographic
change, Social cohesion, Economic vitality and Physical Environment
scores generated by the LLM characterisation and Topic 11 generated by
the structural topic modelling contribute most to between-cluster
differentiation.

\begin{longtable}[]{@{}lr@{}}
\caption{\label{tab:eta_tab} The effect size of each variable on the
clusters.}\tabularnewline
\toprule\noalign{}
Variable & \(\eta^2\) \\
\midrule\noalign{}
\endfirsthead
\toprule\noalign{}
Variable & \(\eta^2\) \\
\midrule\noalign{}
\endhead
\bottomrule\noalign{}
\endlastfoot
Disp & 0.8204 \\
Dem & 0.6810 \\
Soc & 0.4767 \\
Econ & 0.4725 \\
Env & 0.3536 \\
T11 & 0.2654 \\
Diff & 0.1716 \\
T5 & 0.1707 \\
T14 & 0.1644 \\
T7 & 0.1557 \\
T16 & 0.1547 \\
T18 & 0.1136 \\
Gent & 0.1068 \\
T4 & 0.0836 \\
ΔPR & 0.0720 \\
T6 & 0.0376 \\
Decl & 0.0370 \\
T12 & 0.0315 \\
ΔRM & 0.0278 \\
ΔNW & 0.0232 \\
ΔUN & 0.0126 \\
ΔHP & 0.0061 \\
\end{longtable}

\subsection*{GAM summary table}\label{gam-summary-table}
\addcontentsline{toc}{subsection}{GAM summary table}

\begin{longtable}[]{@{}
  >{\raggedright\arraybackslash}p{(\linewidth - 8\tabcolsep) * \real{0.2907}}
  >{\raggedright\arraybackslash}p{(\linewidth - 8\tabcolsep) * \real{0.3140}}
  >{\raggedleft\arraybackslash}p{(\linewidth - 8\tabcolsep) * \real{0.1744}}
  >{\raggedleft\arraybackslash}p{(\linewidth - 8\tabcolsep) * \real{0.1279}}
  >{\raggedleft\arraybackslash}p{(\linewidth - 8\tabcolsep) * \real{0.0930}}@{}}
\caption{\label{tab:gam_tab}GAM summary table.}\tabularnewline
\toprule\noalign{}
\begin{minipage}[b]{\linewidth}\raggedright
Term Type
\end{minipage} & \begin{minipage}[b]{\linewidth}\raggedright
Variable
\end{minipage} & \begin{minipage}[b]{\linewidth}\raggedleft
Estimate / EDF
\end{minipage} & \begin{minipage}[b]{\linewidth}\raggedleft
Std. Error
\end{minipage} & \begin{minipage}[b]{\linewidth}\raggedleft
p-value
\end{minipage} \\
\midrule\noalign{}
\endfirsthead
\toprule\noalign{}
\begin{minipage}[b]{\linewidth}\raggedright
Term Type
\end{minipage} & \begin{minipage}[b]{\linewidth}\raggedright
Variable
\end{minipage} & \begin{minipage}[b]{\linewidth}\raggedleft
Estimate / EDF
\end{minipage} & \begin{minipage}[b]{\linewidth}\raggedleft
Std. Error
\end{minipage} & \begin{minipage}[b]{\linewidth}\raggedleft
p-value
\end{minipage} \\
\midrule\noalign{}
\endhead
\bottomrule\noalign{}
\endlastfoot
Parametric (Coefficient) & Intercept & 4.289 & 0.892 & 0.000 \\
Parametric (Coefficient) & Intercept.1 & 2.003 & 0.839 & 0.017 \\
Parametric (Coefficient) & Intercept.2 & 1.779 & 1.069 & 0.096 \\
Parametric (Coefficient) & Intercept.3 & 2.989 & 0.945 & 0.002 \\
Parametric (Coefficient) & Intercept.4 & 4.473 & 0.891 & 0.000 \\
Smooth (edf) & s.2(X,Y):Intercept & 2.220 & NA & 0.087 \\
Smooth (edf) & s.2(X,Y):economic\_vitality & 6.348 & NA & 0.000 \\
Smooth (edf) & s.2(X,Y):topic\_14 & 1.071 & NA & 0.035 \\
Smooth (edf) & s.2(X,Y):topic\_11 & 1.718 & NA & 0.000 \\
Smooth (edf) & s.2(X,Y):topic\_7 & 1.067 & NA & 0.011 \\
Smooth (edf) & s.2(X,Y):social\_cohesion & 7.268 & NA & 0.000 \\
Smooth (edf) & s.3(X,Y):Intercept & 0.000 & NA & 0.875 \\
Smooth (edf) & s.3(X,Y):economic\_vitality & 0.966 & NA & 0.028 \\
Smooth (edf) & s.3(X,Y):topic\_14 & 2.233 & NA & 0.000 \\
Smooth (edf) & s.3(X,Y):topic\_11 & 2.052 & NA & 0.000 \\
Smooth (edf) & s.3(X,Y):topic\_7 & 0.000 & NA & 0.399 \\
Smooth (edf) & s.3(X,Y):social\_cohesion & 2.229 & NA & 0.000 \\
Smooth (edf) & s.4(X,Y):Intercept & 0.000 & NA & 0.360 \\
Smooth (edf) & s.4(X,Y):economic\_vitality & 6.727 & NA & 0.000 \\
Smooth (edf) & s.4(X,Y):topic\_14 & 1.662 & NA & 0.011 \\
Smooth (edf) & s.4(X,Y):topic\_11 & 1.751 & NA & 0.001 \\
Smooth (edf) & s.4(X,Y):topic\_7 & 0.000 & NA & 0.857 \\
Smooth (edf) & s.4(X,Y):social\_cohesion & 4.666 & NA & 0.000 \\
Smooth (edf) & s.5(X,Y):Intercept & 0.000 & NA & 0.780 \\
Smooth (edf) & s.5(X,Y):economic\_vitality & 4.285 & NA & 0.000 \\
Smooth (edf) & s.5(X,Y):topic\_14 & 2.592 & NA & 0.000 \\
Smooth (edf) & s.5(X,Y):topic\_11 & 0.509 & NA & 0.349 \\
Smooth (edf) & s.5(X,Y):topic\_7 & 1.142 & NA & 0.000 \\
Smooth (edf) & s.5(X,Y):social\_cohesion & 4.537 & NA & 0.000 \\
Smooth (edf) & s.6(X,Y):Intercept & 3.733 & NA & 0.076 \\
Smooth (edf) & s.6(X,Y):economic\_vitality & 6.144 & NA & 0.000 \\
Smooth (edf) & s.6(X,Y):topic\_14 & 1.978 & NA & 0.000 \\
Smooth (edf) & s.6(X,Y):topic\_11 & 2.328 & NA & 0.000 \\
Smooth (edf) & s.6(X,Y):topic\_7 & 1.182 & NA & 0.004 \\
Smooth (edf) & s.6(X,Y):social\_cohesion & 5.759 & NA & 0.000 \\
\end{longtable}

\newpage

\section*{References}\label{references}
\addcontentsline{toc}{section}{References}

\protect\phantomsection\label{refs}

\end{document}